\documentclass[usenatbib]{aastex6}

\usepackage{natbib}
\usepackage{xcolor}
\newcommand{\eps}[1]{\mbox{log~$\epsilon$(#1)}} 
\newcommand\species[2]{#1 {\sc #2}}
\newcommand\iso[2]{$^{\rm #1}$#2}

\def\eg{\mbox{e.g.}}

\def\teff{\mbox{T$_{\rm eff}$}}
\def\logg{\mbox{log~{\it g}}}
\def\vmicro{\mbox{$\xi_{\rm t}$}}

\def\carbiso{\mbox{$^{12}$C/$^{13}$C}}
\def\rpro{\mbox{\textit{$r$}-process}}
\def\spro{\mbox{\textit{$s$}-process}}
\def\ncap{\mbox{\textit{$n$}-capture}}
\def\ffor{\mbox{HIP 54048}}
\def\fsev{\mbox{HIP 57748}}
\def\oneone{\mbox{HIP 114809}}
\def\loggf{\mbox{log~{\it gf}}}
\def\logeps{\mbox{log~$\epsilon$}}

\shorttitle{Near-$IR$ Spectra of Red Horizontal Branch Stars}
\shortauthors{Af{\c s}ar et al.}

\begin{document}

\title{CHEMICAL COMPOSITIONS OF EVOLVED STARS FROM NEAR-INFRARED IGRINS HIGH-RESOLUTION SPECTRA. I. ABUNDANCES IN THREE RED HORIZONTAL BRANCH STARS}

\author{Melike Af{\c s}ar\altaffilmark{1,2},
        Christopher Sneden\altaffilmark{2},
        Michael P. Wood\altaffilmark{3},
        James E. Lawler\altaffilmark{4},
        Zeynep Bozkurt\altaffilmark{1},
        Gamze B{\"o}cek Topcu\altaffilmark{1}, 
        Gregory N. Mace\altaffilmark{2},
        Hwihyun Kim\altaffilmark{5}, and
        Daniel T. Jaffe\altaffilmark{2}
        }

\altaffiltext{1}{Department of Astronomy and Space Sciences,                 
                 Ege University, 35100 Bornova, {\. I}zmir, Turkey;          
                 melike.afsar@ege.edu.tr, \\
                 zeynep.bozkurt@ege.edu.tr,
                 gamzebocek@gmail.com}
\altaffiltext{2}{Department of Astronomy and McDonald Observatory,           
                 The University of Texas, Austin, TX 78712, USA; 
                 afsar@astro.as.utexas.edu, chris@verdi.as.utexas.edu}
\altaffiltext{3}{Department of Physics, University of St. Thomas, 
                 2115 Summit Ave, St. Paul, Minnesota 55104; 
                 mpwood@stthomas.edu}
\altaffiltext{4}{Department of Physics, University of Wisconsin-Madison, 
                 1150 University Ave, Madison, WI 53706; jelawler@wisc.edu}
\altaffiltext{5}{Gemini Observatory, Casilla 603, La Serena, Chile}

\begin{abstract}

We have derived elemental abundances of three field red 
horizontal branch stars using high-resolution ($R$~$\simeq$ 45,000), high 
signal-to-noise ratio (S/N $\gtrsim$ 200) $H$ and $K$ band spectra obtained with 
the Immersion Grating Infrared Spectrograph (IGRINS).
We have determined the abundances of 21 elements including $\alpha$ (Mg, Si, Ca, S), 
odd-Z (Na, Al, P, K), Fe-group (Sc, Ti, Cr, Co, Ni), neutron-capture (Ce, Nd, Yb), 
and CNO group elements. 
S, P and K are determined for the first time in these stars. 
$H$ and $K$ band spectra provide a substantial number of
\species{S}{i} lines, which potentially can lead to a more robust
exploration of the role of sulfur in the cosmochemical evolution of the Galaxy.
We have also derived \carbiso\ ratios from synthetic spectra of the
first overtone (2$-$0) and (3$-$1) $^{12}$CO and (2$-$0) $^{13}$CO lines near 
23440 \AA\ and $^{13}$CO (3$-$1) lines at about 23730 \AA.  
Comparison of our results with the ones obtained from the optical region 
suggests that the IGRINS high-resolution $H$ and $K$ band spectra offer more 
internally self-consistent atomic lines of the same species for several elements, 
especially the $\alpha$ elements. 
This in turn provides more reliable abundances for the elements with analytical 
difficulties in the optical spectral range.

\end{abstract}


\section{INTRODUCTION}\label{intro}

Low mass core helium burning stars appear
in the HR diagram as members of the horizontal branch (HB).
These highly evolved objects have effective temperatures that span
4800~K~$\lesssim$~\teff $\lesssim$ 25,000~K and absolute magnitudes with
only a small range, 0~$\sim$~$M_V$~$\sim$~+1.
HB stars are subdivided into four temperature groups: 
red clump and red horizontal branch (RC and RHB, 
$\sim$4800$-$6000~K),
the RR~Lyrae instability strip ($\sim$6000$-$7500~K),
blue (BHB, $\sim$7500$-$20,000~K), and extreme (EBHB ($\gtrsim$25,000~K).
As discussed in \citealt{afsar18a} 
(hereafter Af{\c s}ar18a), RHB stars are low mass core-helium-burning stars.
Af{\c s}ar18a follows the observational RHB definition given 
by \cite{kaempf05} and defines an RHB range for the stars around solar 
metallicity with 0.5~$\lesssim$~($B-V$)~$\lesssim$~1.0 and 
$-$0.5~$\lesssim~M_{\rm V}$~$\lesssim$~1.5. 
This ``$box$'' also includes so called ``\textit{secondary red clump}'' 
defined by, e.g., \cite{girardi98} and \cite{girardi99}, in which they 
investigate secondary clump for different metallicities in synthetic 
color-magnitude diagrams (e.g. Figure 7 in \citealt{girardi99}).
HB astrophysical interests are many, from distance indicators, to 
stellar population studies, to signposts for interior advanced nucleosynthesis
and envelope mixing.

HB stars with Galactic disk metallicities 
([Fe/H]~$\gtrsim$~$-$1)\footnote{
We adopt the standard spectroscopic notation 
\citep{wallerstein59} that for elements A and B,
[A/B] $\equiv$ log$_{\rm 10}$(N$_{\rm A}$/N$_{\rm B}$)$_{\star}$ $-$
log$_{\rm 10}$(N$_{\rm A}$/N$_{\rm B}$)$_{\odot}$.
We use the definition
\eps{A} $\equiv$ log$_{\rm 10}$(N$_{\rm A}$/N$_{\rm H}$) + 12.0, and
equate metallicity with the stellar [Fe/H] value.}, \eg, metal-rich globular clusters and field thick 
disk stars, almost always occupy the RHB\footnote{
Metal-rich globular clusters with BHB stars are rare;
notable exceptions are the metal-rich globular clusters NGC~6388 and NGC~6441 
([Fe/H]~$\simeq$~$-$0.6) that have both prominent RHB and BHB populations 
\citep{rich97}.}.
In these older populations there is no evolutionary state ambiguity, since
subgiant stars in the RHB temperature domain have much lower luminosities,
$M_{\rm V}$~$\gtrsim$~+3.5.
However in the general Galactic field the low mass thick disk He-burning
stars can share the same (\teff, \logg) or ($B-V$, $M_{\rm V}$) domain with higher-mass
thin disk subgiants, and also can be confused with the occasional thin-disk
He core-burning stars near the high-mass edge of the metal-rich red giant
``clump''.
These stellar population confusions, combined with 
difficulties in determining accurate distances, have hampered the
identification of field 
stars likely to be true RHB stars, thus rendering them under-studied 
compared to red giants.

\cite{kaempf05} conducted a large-scale photometric survey 
of field stars with Hipparcos parallaxes \citep{vanleeuwen07}, and 
proposed a ($B-V$, $M_{\rm V}$) area likely to be dominated by RHB stars.
Early spectroscopic studies of field RHB chemical compositions were 
conducted by \cite{tautvaisiene96, tautvaisiene97} 
and \cite{tautvaisiene01}.
Those studies concentrated on thick disk and halo stars.
\cite{afsar12} (hereafter ASF12) conducted a small-sample (76 star) survey of bright field 
stars with temperatures and luminosities consistent with RHB classification,
identifying 18 probable true RHB stars, 13 of which appeared to be thin-disk
solar metallicity (thus relatively young) objects.

Expanding on this unexpected result, we have recently reported an optical 
high-resolution spectroscopic survey of 340 field RHB candidates
Af{\c s}ar18a.
This study used equivalent width (EW) measurements to derive model 
atmosphere parameters, metallicities, and some $\alpha$-group and Fe-group 
abundance ratios. In Af{\c s}ar18a, kinematics were computed for almost the entire sample from 
data in the Hipparcos and Gaia DR1 \citep{GAIA16} catalogs.
Based only on these data, we estimate that about 150 of the 
Af{\c s}ar18a sample are true RHB stars, and that there is an admixture of 
thin and thick-disk members in the sample.
In \cite{afsar18b} we will present abundances of 
elements requiring synthetic spectrum computations: odd-Z, Fe-group, 
neutron capture ($n$-capture) elements, and the light LiCNO group.
We will also be able to refine the stellar evolutionary and Galactic 
population breakdown of this sample.

Optical high-resolution spectra are useful for stellar 
metallicities and relative abundance ratios of many elements, but 
they have limitations particularly among the lighter elements.
Such deficiencies can be overcome by obtaining data in the infrared ($IR$)
spectral domain, $\lambda$~$>$~1~$\mu$m.
To this end we have initiated a program to observe a large subset of
the Af{\c s}ar18a sample in the photometric $H$ and $K$ bands with the
Immersion Grating Infrared Spectrometer (IGRINS).
With these spectra we will be able to study light elements 
either not available or poorly studied in the optical 
spectral range (P, S, and K),
greatly refine the abundances of
those with some transitions in both the optical and IR regions 
($\alpha$ elements Mg, Si, and Ca; odd-Z elements 
Na and Al; $n$-capture elements Ce, Nd, Yb), 
and most importantly improve the abundances of CNO and the carbon 
isotopic ratios derived from our optical data.

In this paper, we present IGRINS observations of three RHB stars from 
Af{\c s}ar18a (also previously investigated in ASF12): \ffor, \fsev\ and \oneone.
We discuss atomic line identifications in the IGRINS data and the transition
parameters of the chosen lines.
A detailed comparison is made between the resulting abundances and those 
derived from the optical spectra.
A future paper will contain the results from the full IGRINS survey, about
70 stars.
In \S\ref{observ} the IGRINS observations and their
reductions are described. In \S\ref{model} we present the model atmosphere
parameters and abundances of elements derived from the optical spectral region.
Abundance determinations from the IGRINS $H$ and $K$ band spectra are
described in \S\ref{abund} along with the atomic and molecular data used in
abundance determinations. In \S\ref{fei} we investigate the infrared 
\species{Fe}{i} lines as temperature indicators. 
Finally, we discuss and summarize our results in \S\ref{disc}.

\section{OBSERVATIONS AND DATA REDUCTION}\label{observ}

High-resolution, high signal-to-noise ($S/N$~$>$~200 per resolution 
element) spectra of \ffor, \fsev\ and \oneone\ were obtained 
with IGRINS on the 2.7m Harlan J. Smith Telescope (HJS) at McDonald 
Observatory. 
The data were gathered in 2014 during an instrument 
commissioning run on May 24 and 27, and then on a separate run on October 19.
We summarize the basic parameters and observation log in Table~\ref{tab-basic}.

\begin{figure}
\epsscale{0.75}
\plotone{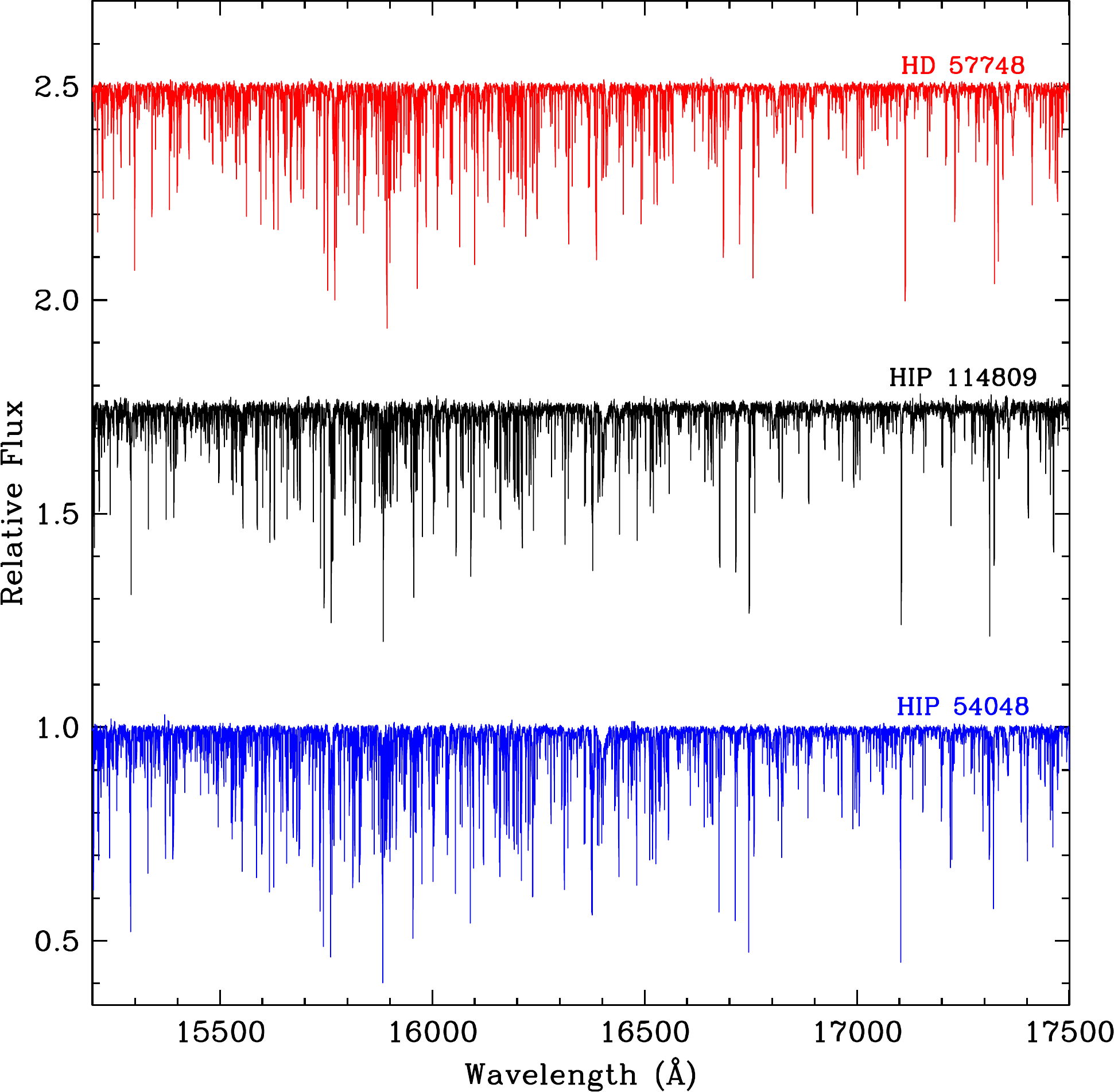}
\caption{
\label{hspec}
\footnotesize
IGRINS $H$ band spectra of three RHB stars in air wavelengths: HIP~57748 (red),  
HIP~114809 (black), and
HIP~54048 (blue). 
For plotting clarity, we shifted the relative flux scale of HIP~57748 and HIP~114809 
vertically with additive constants .
The low and high wavelength edges of the spectra have been trimmed to 
avoid the edges with severe telluric contamination.
}
\end{figure}

\begin{figure}
\epsscale{0.75}
\plotone{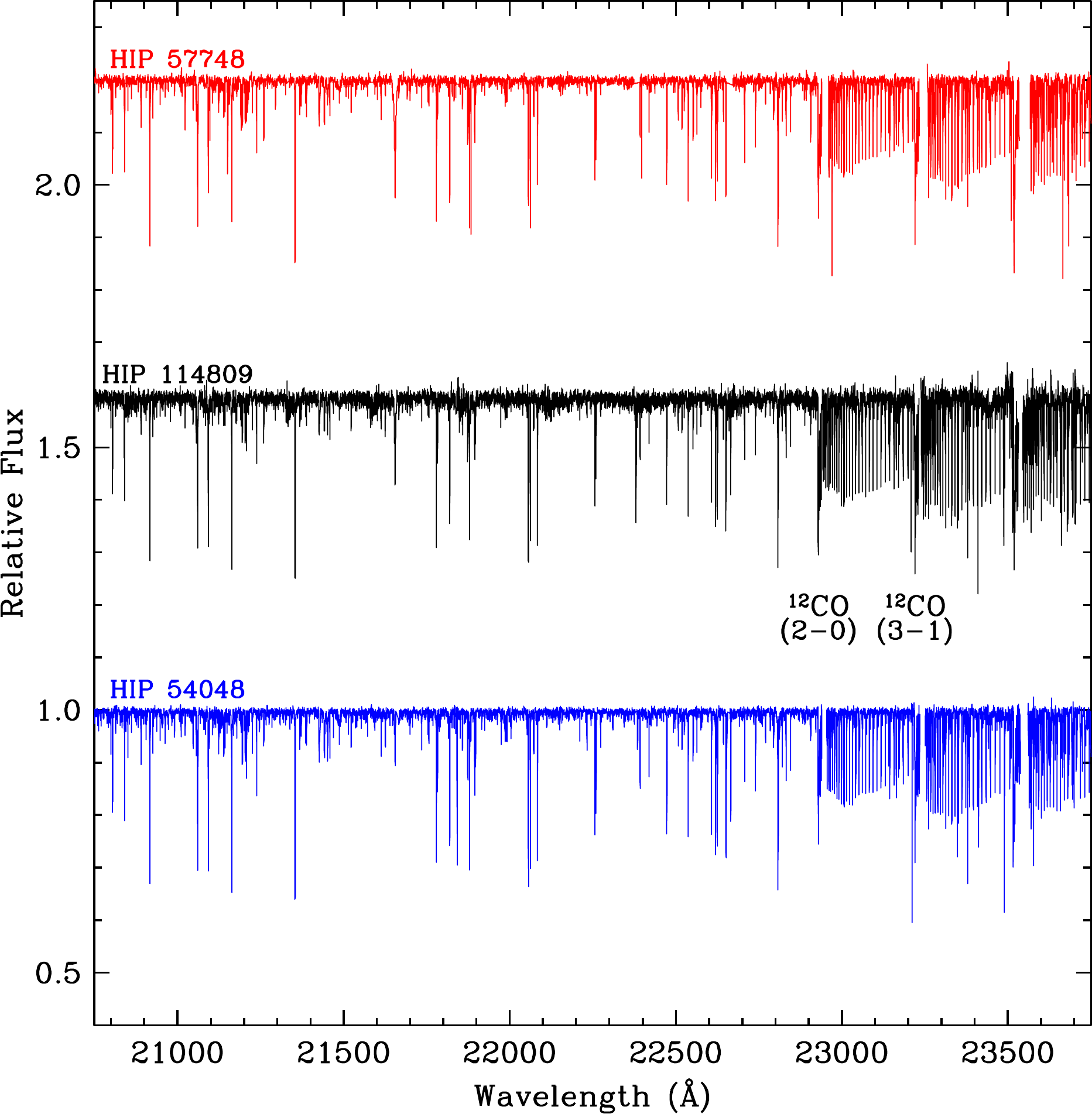}
\caption{
\label{kspec}
\footnotesize
IGRINS $K$ band spectra of three RHB stars in air wavelengths: HIP~57748 (red),  
HIP~114809 (black), and
HIP~54048 (blue). The loci of the $^{12}$CO (2$-$0) and $^{12}$CO (3$-$1)
band heads are also shown. 
}
\end{figure}

\begin{figure}
\epsscale{0.95}
\plotone{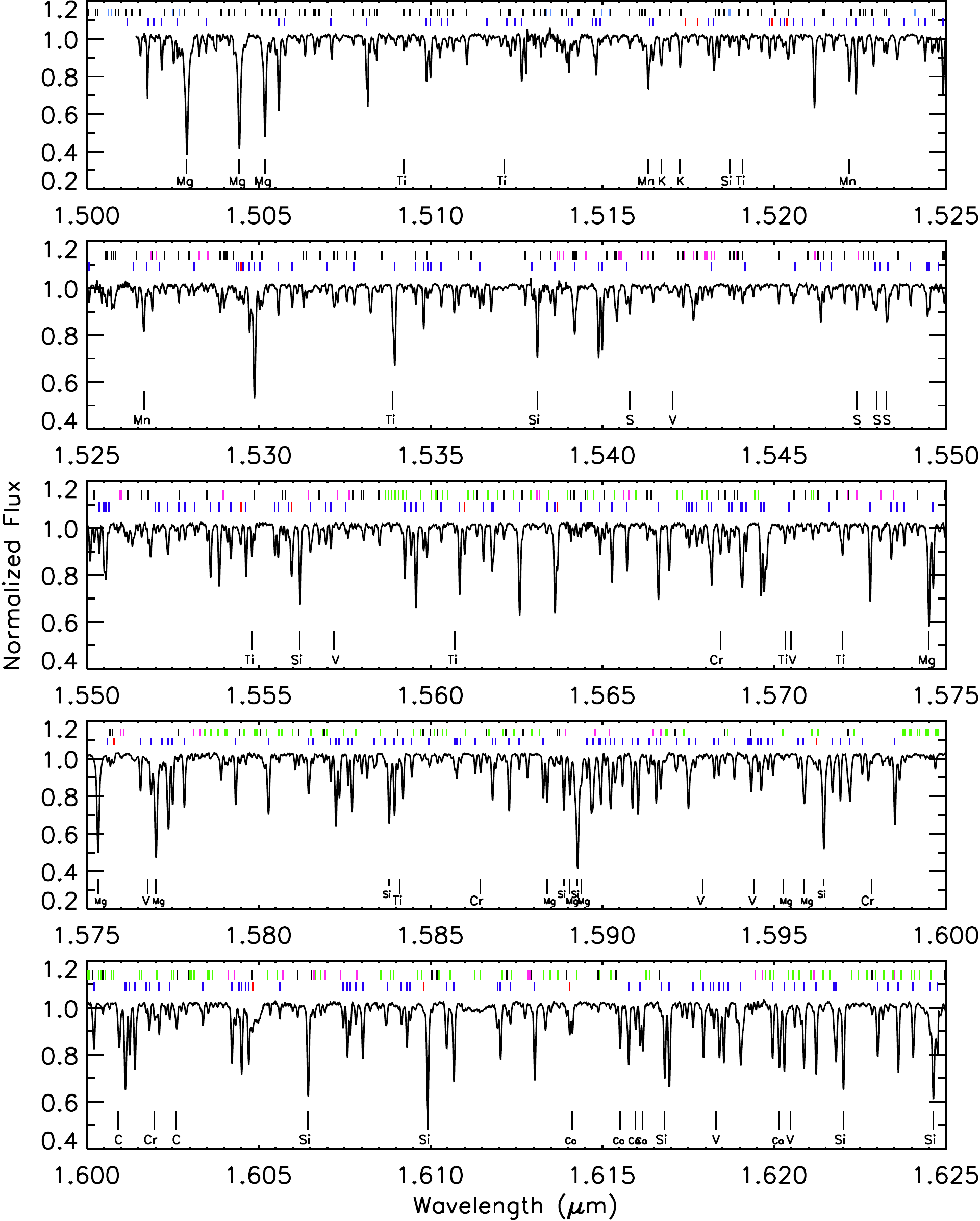}
\caption{
\label{atlas}
\footnotesize
A portion of the $H$ and $K$ band spectral atlas for our stars is presented.
We use the \ffor\ spectrum as the benchmark of RHB stars. In this figure the atomic
lines are labeled with their names, including substantial amount of 
Fe and Ni, which are indicated with blue and red ticks, respectively,
on the top second row of the figure panel.  
CN, OH and CO molecular features are illustrated with black, magenta and green 
ticks, respectively, on the top row of the figure panel.
All line identifications are vacuum wavelengths (\citealt{hinkle95}). 
The IGRINS spectrum has been shifted to the reference frame of the line lists.
}
\end{figure}

An important capability of IGRINS is that it obtains the complete
coverage of $H$ and $K$ bands at high resolving power
($R$~$\equiv$ $\lambda/\Delta\lambda$~$\simeq$ 45,000) simultaneously without
any adjustments to the instrument components.
In a single exposure,
the data cover a wavelength range between 14,800$-$24,800~\AA, 
with a small gap of about 100~\AA\ between bands. 
More detailed and extended descriptions of IGRINS are given in \cite{yuk10}
and \cite{park14}. 
A discussion of our observational techniques and
reduction processes is given in \cite{afsar16} (hereafter Af{\c s}ar16).
Figure~\ref{hspec} and Figure~\ref{kspec} shows the ready-for-analysis 
spectra of
three RHB stars: wavelength-calibrated, continuum-normalized, 
telluric-line-removed, and merged into a single continuous spectrum.
We have restricted the figures in both $H$ and $K$ band wavelengths so 
that the heavily sky-contaminated band edges that were not used 
in the analysis are not shown.
We also present a detailed spectral atlas for our stars taking \ffor\ as the 
benchmark of RHB stars. A portion of the atlas is illustrated in Figure~\ref{atlas} (the complete 
figure is available online).

The optical spectra of \ffor, \fsev\ and \oneone\ were 
obtained during observing runs in 2009 and 2010, and their analyses were
first presented in ASF12 and refined in Af{\c s}ar18a.
The details of the optical observations and data reduction are 
described in that paper.

\section{MODEL ATMOSPHERIC PARAMETERS AND ABUNDANCES FROM OPTICAL-REGION SPECTRA}\label{model}

Stellar atmospheric parameters of our program stars have 
been recently revisited by Af{\c s}ar18a. 
Highlights of the new analyses include: (a) revision of the model atmospheric 
parameters by including the neutral and ionized species of Ti lines in the 
calculations along with \species{Fe}{i} and \species{Fe}{ii} lines, 
and (b) updates to the line oscillator strengths from recent laboratory
work.
Af{\c s}ar18a derived \teff, \logg, [Fe/H], and \vmicro\ using a 
semi-automated iterative driver operating on the current version of the local
thermodynamic equilibrium (LTE) line analysis and synthetic spectrum 
code MOOG \citep{sneden73}\footnote{
Available at http://www.as.utexas.edu/$\sim$chris/moog.html}.
In this study we adopt these revised model atmospheric parameters; they
are listed in Table~\ref{tab-model}.

Atomic and molecular lines that we used to derive both atmospheric 
parameters and individual element abundances 
from the optical spectra
are as described in \cite{bocek15} and \cite{bocek16}. 
The line list used in this study is given in 
Table~\ref{tab-lines} along 
with oscillator strength information and references.

In total, we derived the elemental abundances of 21 species of 18 elements 
in the optical region by using EW measurements and 
spectrum syntheses (SYN) for those lines with hyperfine (HFS) structures and/or 
blending features. The abundances of elements Si, Ca, Ti, Cr, and Ni 
that were determined from EWs are taken from Af{\c s}ar18a; details about
those calculations are given below in \S\ref{alphaodd-opt}.
To find the differential abundances relative to the Sun, we adopted solar 
abundances from \cite{asplund09}. 
The individual abundances are given in Table~\ref{tab-lines}, and average 
[X/Fe] values along with standard deviations and numbers of lines for all species
are summarized in Table~\ref{tab-54048}, \ref{tab-57748} and 
\ref{tab-114809}.

\subsection{$\alpha$ and Odd-Z Light Elements}\label{alphaodd-opt}

We determined the abundances of $\alpha$ elements Mg, Si, S 
and Ca from their neutral species transitions. 
Abundances of Si and Ca were derived from EW measurements, 
while SYN was applied to determine Mg and S abundances.

There are only a few useful \species{Mg}{i} 
lines in the optical spectral region.
Mg abundances in our stars were determined from 
two strong \species{Mg}{i} lines at 5528.4 and 5711.1 \AA.
They are slightly overabundant in all three stars, with little star-to-star 
scatter: $\langle$[Mg/Fe]$\rangle$~=~$+$0.16.
Considering the large strength of these transitions, their
derived abundances are very dependent on microturbulent velocities and
other outer-atmosphere line formation effects.

Si is significantly overabundant in all three RHB stars,
$\langle$[Si/Fe]$\rangle$~=~$+$0.24.
There are 13-14 \species{Si}{i} lines participating in the analysis of each 
star, so statistically the Si abundance should be well determined.
However, in Af{\c s}ar18a we cautioned that the derived Si 
abundances are nearly always high, even in solar-metallicity stars. 
Moreover, there is a relatively large star-to-star scatter,
and a significant trend with \teff; see Af{\c s}ar18a Figure~13.
We recommend caution in interpreting these Si abundances.

Optical S lines are rarely studied. 
We derived the S abundances for our RHB stars using the triplet 
centered at 6757.0 \AA.
\cite{takeda16} investigated non-local thermodynamic equilibrium (nLTE) 
effects for this triplet in G-K giant stars, finding corrections to 
be mostly $\lesssim$0.1 dex, in good agreement with earlier computations 
\citep{korotin09}. Sulfur abundances in our stars 
are essentially solar, $\langle$[S/Fe]$\rangle$~=~+0.09.
Other studies have mostly used the \species{S}{i} 
transitions at 8693$-$5 \AA.
We could not study these lines because they fall into an echelle order gap in 
our spectra.

There are many \species{Ca}{i} lines in the optical 
wavelength region, and Af{\c s}ar18a uses 11 of them.
The resulting abundance is slightly above solar, 
$\langle$[Ca/Fe]$\rangle$~=~$+$0.12.
In total, Mg, Si, S, and Ca all indicate [$\alpha$/Fe]~$\simeq$~$+$0.15, 
consistent with overall [$\alpha$/Fe] levels in these mildly metal-poor 
disk stars (see Figure~14 of Af{\c s}ar18a).

We derived the abundances of three light odd-Z elements, Na, Al, and K 
using the SYN method.
Na abundances were usually determined from four \species{Na}{i} lines 
located at 5682.6, 5688.2, 6154.2 and 6160.6 \AA. 
As is known from several studies (e.g. 
\citealt{takeda03,lind11}), these lines are subject to nLTE effects.
\cite{takeda03} suggest that the \species{Na}{i} lines at 6154.2 and 
6160.8~\AA\ are the least sensitive to the nLTE effects, and the 
corrections needed for stars like our RHB stars are less than
 $-$0.1 dex, while the lines at 
5682.6 and 5688.2~\AA\ need nLTE corrections up to about $-$0.15 dex.
These results are also in 
good agreement with \cite{lind11}, who studied the nLTE
calculations for neutral Na lines in late-type stars.
With these recommended nLTE corrections, 
all four \species{Na}{i} lines yielded internally consistent abundances
for each of the program stars.
However, there is large star-to-star scatter, with HIP~54048, HIP~57748,
and HIP~114809 giving abundances [Na/Fe]~= $+$0.25, $+$0.03, and +0.34,
respectively.
 
For Al abundances we used neutral species transitions
at 6696.0, 6698.7, 7835.3 and 7836.1 \AA. 
Recently \cite{nord17} have investigated nLTE effects on several 
\species{Al}{i} lines, including the ones we report in this study. 
They conclude that the nLTE corrections needed for the 
\species{Al}{i} doublets at $\lambda$$\lambda$6696, 6698 and 
$\lambda$$\lambda$7835, 7836 are not more than $-$0.05 dex for cool 
giants with solar metallicities.
As in the case of Na, the internal line-to-line scatter is small for 
the four \species{Al}{i} transitions but the star-to-star scatter is
substantial:  [Al/Fe]~=~$+$0.05, $-$0.15, and +0.10 for the three stars
listed as above.
Thus it appears that HIP~57748 has 0.2-0.3~dex smaller Na and Al 
abundances than do the other two stars.

The only available optical \species{K}{i} lines are 
those of the resonance doublet at 7665 and 7699~\AA.
Usually the 7665~\AA\ line is too blended with the telluric A band
to be of any use in abundance analyses. 
These very strong lines are severely affected by uncertainties in 
microturbulent velocities, adopted damping constants, and large
nLTE effects (\citealt{takeda02}). 
Investigation of the \species{K}{i} 7699~\AA\ line in our stars yielded 
$\sim$0.6 dex higher abundances in all RHB stars compared to that expected for 
disk stars ($-$1.0 $\lesssim$ [Fe/H] $\lesssim$ 0.0)  
(see Figure~4 in \citealt{takeda02} and Figure~6 in \citealt{takeda09}).  
Those two papers report nLTE corrections ($\Delta_{\rm nLTE}$) up to 
$-0.7$ dex.
In a more recent study, \cite{mucciarelli17} also 
suggest $\Delta_{\rm nLTE}$ corrections up to $\sim -$0.6 dex.
Applying these large nLTE corrections bring our derived K abundances into
reasonable agreement with the solar values.

\subsection{Fe-group and neutron-capture elements}\label{fencap-opt}

The Fe-group elements investigated in this study are
Sc, Ti\footnote{
Ti is often classified as an $\alpha$ element.
The major naturally-occurring isotopes of true $\alpha$ elements 
Mg, Si, S, and Ca are composed of multiples of \iso{4}{He} nuclei. 
However, the Ti abundance is dominated by \iso{48}{Ti} (73\% in the 
solar system), and the $\alpha$ isotope \iso{44}{Ti} is unstable with a 
short half-life. Ti is better classified as a Fe-peak element.}, Cr, Co, and Ni.
Abundances from \species{Ti}{i}, \species{Ti}{ii}, \species{Cr}{i}, 
\species{Cr}{ii}, and \species{Ni}{i} transitions were determined in Af{\c s}ar18a
using EW matching.
Ti abundances from both neutral and ionized species were also 
used to help derivations of model atmospheric parameters. 
Sc and Co abundances were not reported in Af{\c s}ar18a because \species{Sc}{ii} 
and \species{Co}{i} lines have significant HFS substructure, and 
abundance extractions from them need special treatment. 
For these elements here we have used
MOOG's \textit{blends}
option that produces EWs from complex spectral features.  
Recent lab data for \species{Co}{i} are taken from \cite{lawler15}.  
For \species{Sc}{ii} lines we adopted the lab \loggf\ values 
from \cite{lawler89}.
Abundances of the Fe-group elements appear to be mostly solar in all 
RHB stars, with the exception of modest deficiencies observed in Ti and Sc in \fsev,
and a slight Sc overabundance in \oneone.

In this study, we have only investigated the 
neutron-capture (\ncap) elements Ce and Nd.
Some other \ncap\ elements present in the optical spectra 
of these stars will be reported in the next survey paper \cite{afsar18b}.
Ce and Nd are two \ncap\ elements whose solar system abundances are
due mostly to relatively slow \ncap\ events (the \spro).
We determined the optical Ce abundances from four \species{Ce}{ii} lines. 
The lab \loggf\ data are taken from \cite{lawler09}. 
Nd abundances were measured from two \species{Nd}{ii} lines at 5255.5 
and 5319.8 \AA. 
Their lab data were gathered from \cite{denhartog03}. 
The abundances determined from the SYN method of both
[Ce/Fe] and [Nd/Fe] vary from star to star in our small sample:
slightly overabundant in \ffor, around solar in \oneone,
and more than $-$0.2 dex deficient in \fsev.

\subsection{The CNO Group}\label{light-opt}

We determined CNO abundances by application of the
SYN method to complex atomic and molecular features.
To derive C abundances we analyzed CH G-band 
Q-branch band head lines in the 4300$-$4330 \AA\ region and 
lines of the C$_{2}$ Swan band heads near 5160 and 5631 \AA.

C abundances can be also derived from \species{C}{i} high 
excitation potential ($\chi$~$\gtrsim$~7.7 eV) lines, and ASF12
attempted to apply $EW$ analyses of them
 for their initial sample of RHB stars. 
The drawbacks of these lines include rapid decrease in strength
with decreasing \teff\ and probable nLTE effects. 
Here we determined C abundances 
from SYN analyses of three high-excitation \species{C}{i} 
lines that have been used in previous
studies (e.g. \citealt{caffau10})
located at 5052.1, 5380.3 and 8335.1 \AA. 
We list the individual line by line  \logeps\ values in Table~\ref{tab-c}.
The line-to-line scatter among the \species{C}{i} lines remains
within $\sim$0.1 dex.

For N abundances we analyzed the 7995$-$8040 \AA\ region that contains
$^{12}$CN and $^{13}$CN red system lines. 
The same region was also used to derive carbon isotopic ratios (\carbiso).
O abundances were derived from two forbidden lines: [\species{O}{i}]
6300.3 \AA\ and [\species{O}{i}] 6363.8 \AA. 

CNO abundances of our targets were previously investigated by ASF12.
The results we report in this study are updated abundances 
(Table~\ref{tab-54048}, \ref{tab-57748} and \ref{tab-114809}) 
using the revised model atmosphere parameters from Af{\c s}ar18a 
(Table~\ref{tab-model}) and different solar abundances from those of ASF12.
Here we are using the \cite{asplund09} recommended CNO abundances
\logeps(C)$_\odot$~=~8.43, \logeps(N)$_\odot$~=~7.83, and
\logeps(O)$_\odot$~=~8.69, 
instead of the ASF12 values of 8.38, 8.2, and 8.64, respectively.
Comparison of these two solar CNO abundance sets is difficult because
\citeauthor{asplund09} used several abundance indicators for each element and 
a variety of analytical approaches, while ASF12 concentrated on CH, CN, and
[\species{O}{i}] using SYN as in the present work.
A closer comparison would be the lab/solar/stellar study of \cite{sneden14},
who derived \logeps(C)$_\odot$~=~8.48, \logeps(N)$_\odot$~=~8.05, and
\logeps(O)$_\odot$~=~8.69, which essentially agrees with \citeauthor{asplund09}
for C and O, but has a 0.2~dex larger N abundance.
The difference in adopted solar N abundance is larger between ASF12 and
\citeauthor{asplund09}:
$\Delta$\logeps(N)~=~\logeps(N$_{\odot}$)$_{\rm{ASF12}}$$-$\logeps(N$_{\odot}$)$_{\rm{Asplund}}$~$\simeq$~0.4 dex, which leads naturally in a large difference
in our [N/Fe] values compared to those of ASF12.
Lower metallicity values ($-$0.13 dex on average) in the Af{\c s}ar18a revised 
model atmospheres of all RHB stars also contribute to [X/Fe] offsets for the CNO
group.
The reader should keep these differences in mind in interpretations of these
elemental abundances.

Scale differences between the present work and 
that of ASF12 do not change the general conclusions about the CNO group.
Their abundances follow the general trends of evolved disk giants:
C abundances decrease, while N abundances increase
due to convection movement of the CN-processed material from inside out 
to the stellar surface, and O holds its own essentially solar value.

\begin{figure}
\epsscale{0.95}
\plotone{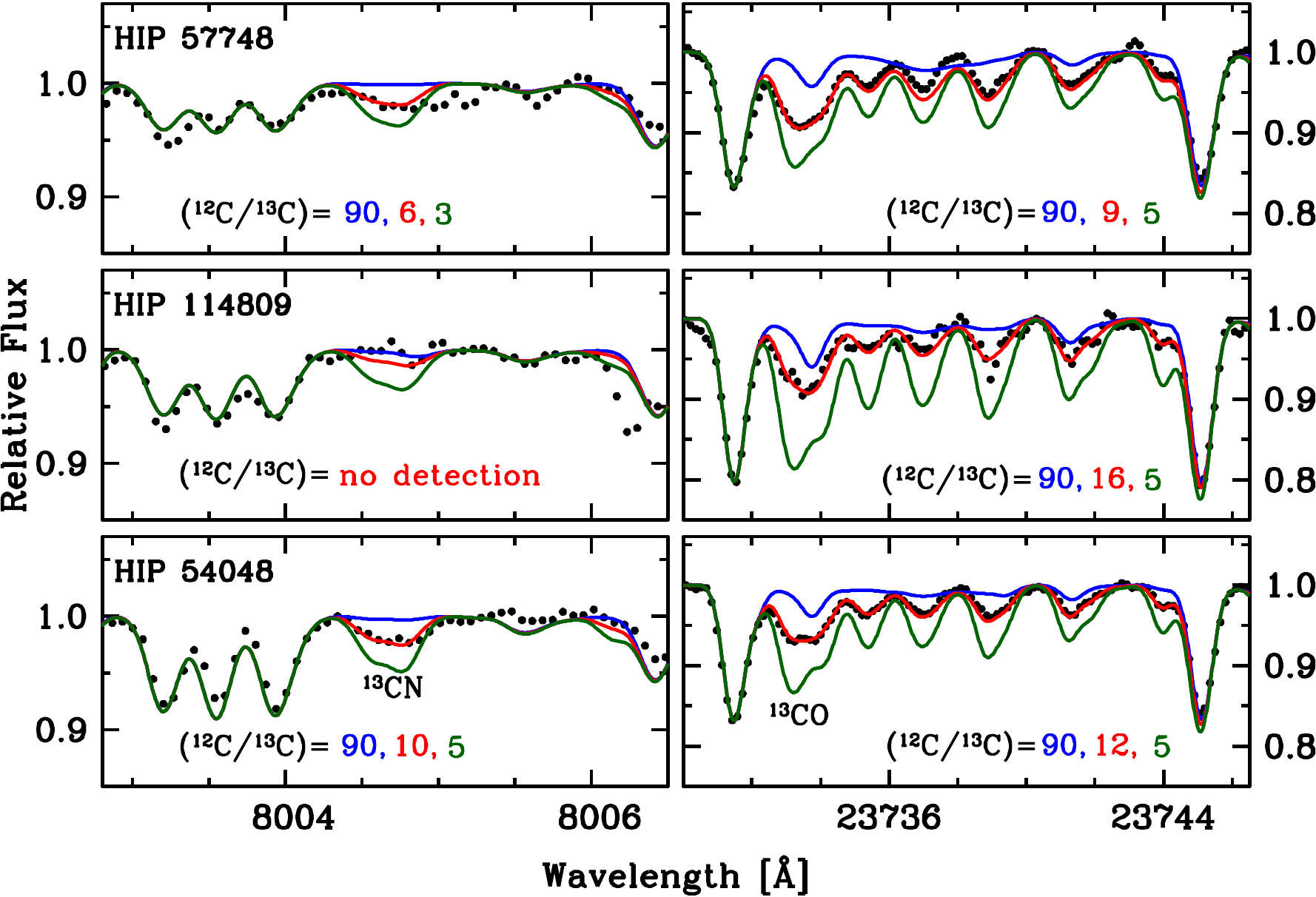}
\caption{
\label{carbiso}
\footnotesize
$^{13}$CN at 8004.7 \AA\ and $^{13}$CO (3$-$1) band head in the $K$ band
are shown for our targets.
Observed spectra are given in black dots. Blue (top), red (middle) and green (bottom) 
lines represent the synthetic spectra for assumed \carbiso~ ratios. 
The red line is the best matching synthesis to the observed spectrum. 
Assumed \carbiso~ ratios are also given in the figure panels. 
}
\end{figure}

We also determined carbon isotopic ratios from the 
primary \carbiso\ indicator near 8004.7 \AA, which is a triplet of $^{13}$CN lines.
In the left panels of Figure~\ref{carbiso} we illustrate the weakness
of $^{13}$CN absorption in our relatively warm RHB stars.
As previously reported in ASF12, a meaningful value of \carbiso\ 
of \oneone\ in the optical region could not be derived.
We list the derived ratios in Table~\ref{tab-carbiso}; these
will be discussed further in \S\ref{cno-ir}.

\section{ABUNDANCES FROM IGRINS NEAR-INFRARED SPECTRA}\label{abund}

High-resolution, high signal-to-noise $H$ and $K$ band 
spectra of RHB stars reveal a rich assortment of atomic and molecular
features (Figure~\ref{atlas}). 

Having this opportunity, we were able to compute the abundances
of 21 species of 20 elements along with \carbiso\ ratios.
For our three RHB stars, abundances of P, S, and K were obtained 
for the first time. 
All abundances and \carbiso\ ratios
were derived by applying SYN method to the observed spectra.
The atomic and molecular data used for abundance calculations and the
abundance analysis of all species are discussed in the following subsections.  
In all cases the individual line abundances are given in 
Table~\ref{tab-lines} and the abundance summaries for the stars are given 
in Tables~\ref{tab-54048}, \ref{tab-57748}, and \ref{tab-114809}.

\subsection{Atomic and molecular data}\label{atomic}

We adopted the molecular data described in 
Af{\c s}ar16 (and references therein) for CN, CO, and OH features,
and generated the atomic line list (Table~\ref{tab-lines}) following a 
similar procedure to that of Af{\c s}ar16. 
We used the Arcturus infrared atlas \citep{hinkle95}\footnote{
Available at ftp://ftp.noao.edu/catalogs/arcturusatlas/}
along with atomic line lists from the Nationl Institute of Standards and
Technology (NIST) Atomic Spectra Database\footnote{
https://physics.nist.gov/PhysRefData/ASD/lines\_form.html}
and from the \cite{kurucz11} database\footnote{
http://kurucz.harvard.edu/linelists.html.}
to identify promising atomic lines in our RHB stars. 
Accurate transition probability information for the atomic and molecular 
data is essential for a robust abundance analysis. 
For lines that have NIST transition probabilities rated A and B accuracy
(uncertainty $\leq$10\%), we adopted the NIST \loggf\ values. 
However, laboratory transition 
probabilities for most of the species transitions in the $IR$ are lacking. 
Therefore, we had to apply reverse solar analyses to determine the
``astrophysical'' \loggf\ values of mostly neutral-species of elements 
listed in Table~\ref{tab-lines}. 
Only a few ionized-species could be identified (see \S\ref{fencap-opt}), 
and among them we only applied reverse solar analysis to \species{Yb}{ii}. 
The transition probabilities ranked as C (accuracies $<$ 20\%) in the NIST 
database were needed to be justified in most of the cases 
with their own reverse solar analyses.
The sources of all transition probabilities are provided in 
Table~\ref{tab-lines}.

For the reverse solar analyses we used the 
high-resolution infrared solar flux spectrum of \cite{wallace03}, and
adopted the solar model atmosphere of ASF12. 
The derived transition probabilities then were tested on the Arcturus atlas
spectra \citep{hinkle95}, adopting the Arcturus abundances and model 
atmospheric parameters from \cite{ramirez11}.
Satisfactory agreement with the abundances derived by \citeauthor{ramirez11} 
was achieved for all the elements investigated in this manner.

Applying reverse solar analysis to obtain astrophysical oscillator 
strengths comes with its own handicaps. 
\cite{barklem00} computed collisional broadening 
cross-sections for many neutral species transitions in $UV$ and optical spectral
regions, but their work stops short of the $H$ and $K$ band regions. 
Broadening cross-sections of \species{Fe}{ii} lines in 
the $H$ and $K$ bands 
have been provided by \cite{barklem05}, but these transitions are too weak
to be detected for our RHB IGRINS data.
Therefore, we had to resort initially to the classical \cite{unsold55}
approximation, which is known to underestimate the observed
line broadening due to van der Waals interactions (e.g. \citealt{chen00}). 
This approximation especially has difficulties in transitions that occur 
at higher-excitation potentials ($\chi$$_{\rm low}$~$>$~3.0 eV, \citealt{chen00}, and references therein), 
which is mostly the case in our lines.
Thus, we applied a damping enhancement factor of about 3.2 
on average to the van der Waals damping computed according to 
the Uns{\"o}ld approximation in order to better estimate the 
collisional broadening effect on absorption lines.
Although applying enhancement factors temporarily cure the problem 
in an empirical way, the underlying physics of the strong broadening we 
observe in the lines of the infrared solar spectrum should be investigated in the future.

\begin{figure}
\epsscale{0.75}
\plotone{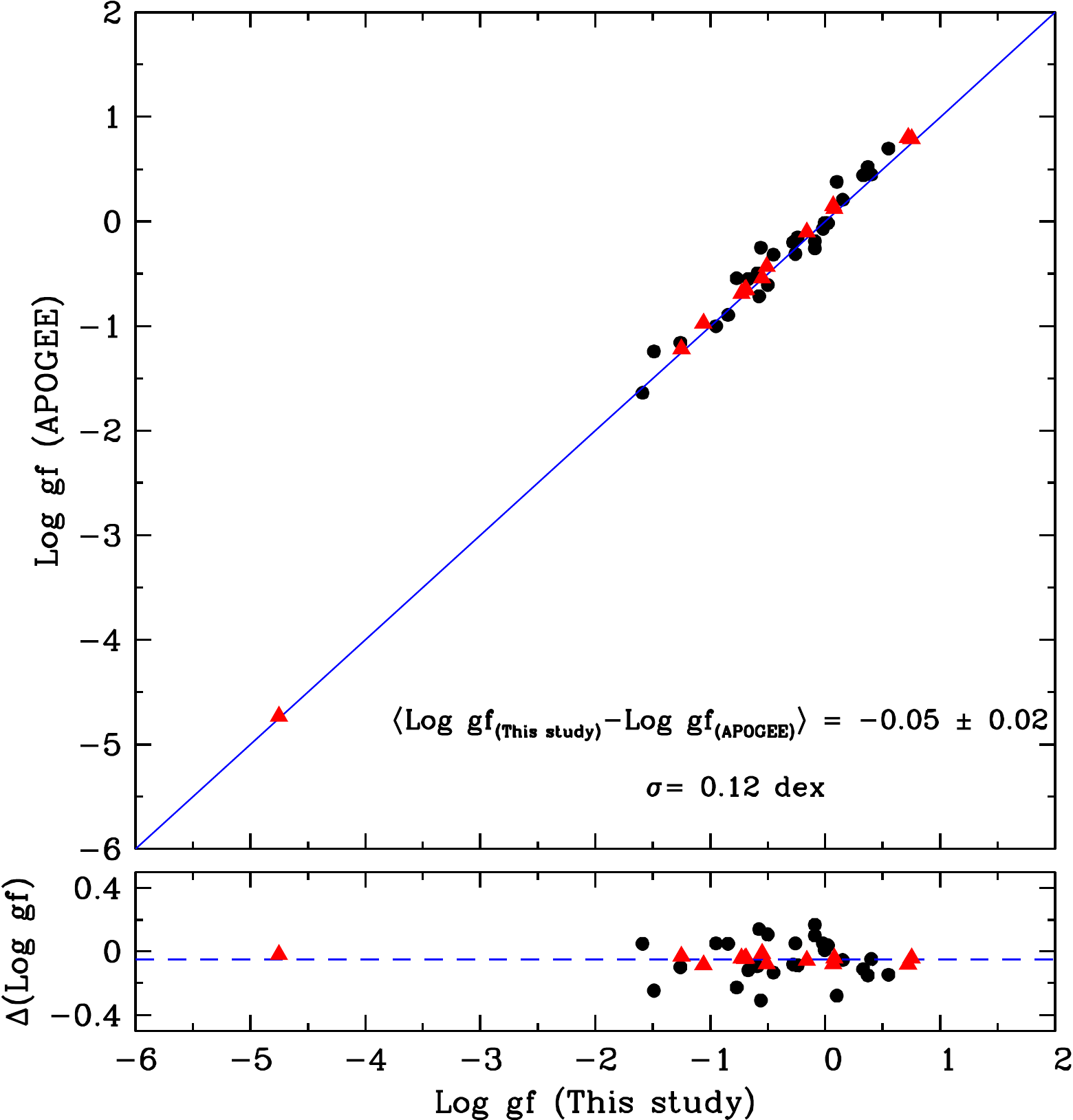}
\caption{
\label{Fe_loggf}
\footnotesize
Comparison of \loggf\ values (This study vs. APOGEE) of nine different
elements (see \S\ref{atomic}) are plotted in the top panel. 
Red triangles represent the \loggf\ values of  \species{Fe}{i} lines. Bottom panel
shows the $\Delta$(\loggf) = \loggf$_{\rm (This~study)}$$-$\loggf$_{\rm (APOGEE)}$.
In the top panel, the blue line denotes equality of the \loggf\ values. The blue dashed 
line in the bottom panel denotes the mean of the differences. 
 }
\end{figure}

We compare our astrophysical oscillator strengths with 
the ones reported by \cite{shetrone15} for the APOGEE high resolution $H$-band
spectroscopic survey (\citealt{majewski16,majewski17}\footnote{
http://www.sdss.org/dr14/irspec/\#APOGEESurveyandInstrumentsOverview})
for the 40 lines in common in both studies in Figure~\ref{Fe_loggf}:
13 Fe, two Ti, seven Si, two P, three S, two K, four Ca, two Cr and five Ni 
(all neutral species transitions).
As seen in the figure, the astrophysical oscillator strengths from both 
studies are mostly in good agreement within $\sigma$ = 0.12 dex. 
We also compared our damping parameters with the ones reported by 
\cite{shetrone15}, in which they also predict enhanced damping for most of the $IR$
lines in common.  
They define the damping parameter as:
vdW = log ($\Gamma$$_{6}$/$N$$_{\rm H}$); $\Gamma$$_{6}$ is the 
van der Waals collisional damping and $N$$_{\rm H}$ is the 
number density of hydrogen.
Comparison of our vdW values with \citeauthor{shetrone15} resulted in
reasonable agreement,
$\langle$vdW$\rangle$ = $-$0.02 ($\sigma$ = 0.25).

\subsubsection{Isotope Shift in \species{Ti}{i}}\label{tiso}

To the best of our knowledge, \species{Ti}{i} lines 
are the only ones whose IR profiles have been investigated in detail.
\cite{blackwell06} called attention not to damping effects, but to isotopic 
wavelength splits.
They found that \species{Ti}{i} $IR$ lines have significant isotopic 
broadening.
Although most of Ti is in the form of \iso{48}{Ti}, it has five 
naturally-occurring isotopes: \iso{46}{Ti}, 8.25\%; 
\iso{47}{Ti}, 7.44\%; \iso{48}{Ti}, 73.72\%; \iso{49}{Ti}, 5.41\%; 
and \iso{50}{Ti}, 5.18\%.\footnote{
Chang, J.: Table of Nuclides, KAERI(Korea Atomic 
Energy Research Institute).  Available at: 
http://atom.kaeri.re.kr/ton/. Retrieved June 11, 2018}

Isotope shifts (ISs) in \species{Ti}{i}, like other spectra, are due to a 
combination of Normal Mass Shift (NMS), Specific Mass Shift (SMS), and 
Field Shift (FS). 
The NMS is the elementary reduced mass shift of classical or quantum mechanical 
two body systems.
Elements near the top of the periodic table have dominant NMSs 
starting with \species{H}{i} vs \species{D}{i} ISs. 
The SMS arises from terms in the Hamiltonian involving the momentum of 
one electron dotted with the momentum of another electron. 
Multi-electron atoms, including Ti, can have significant SMSs. 
Lastly the FS appears from changes in the mean squared charge radius of 
the nucleus as neutrons are added. 
In general one finds large ISs near the top of the periodic table dominated by 
mass shifts, and large ISs near the bottom of the periodic table dominated 
by FSs especially when there is an s-electron participating in the atomic 
transition.

The important Fe-group spectra tend to have small ISs that are difficult 
to detect in stellar spectra. 
However, this situation may change due to the development of HgCdTe detector 
arrays and the increase in high spectral resolution, high S/N stellar data. 
The study of atomic transitions at lower frequencies, e.g. in the IR, will 
generally yield larger fractional ISs. 
Data in this work on the IR lines of \species{Ti}{i} clearly reveal ISs. 
Studies of pre-solar grains in chondrites have revealed many variations in 
isotopic abundance patterns (e.g. \citealt{lodders05}) that result from 
different nucleo-synthetic histories of grain material. 
The possibility that the IR may enable more detailed studies of isotopic 
abundance patterns using stellar spectra is appealing.

\begin{figure}
\epsscale{0.75}
\plotone{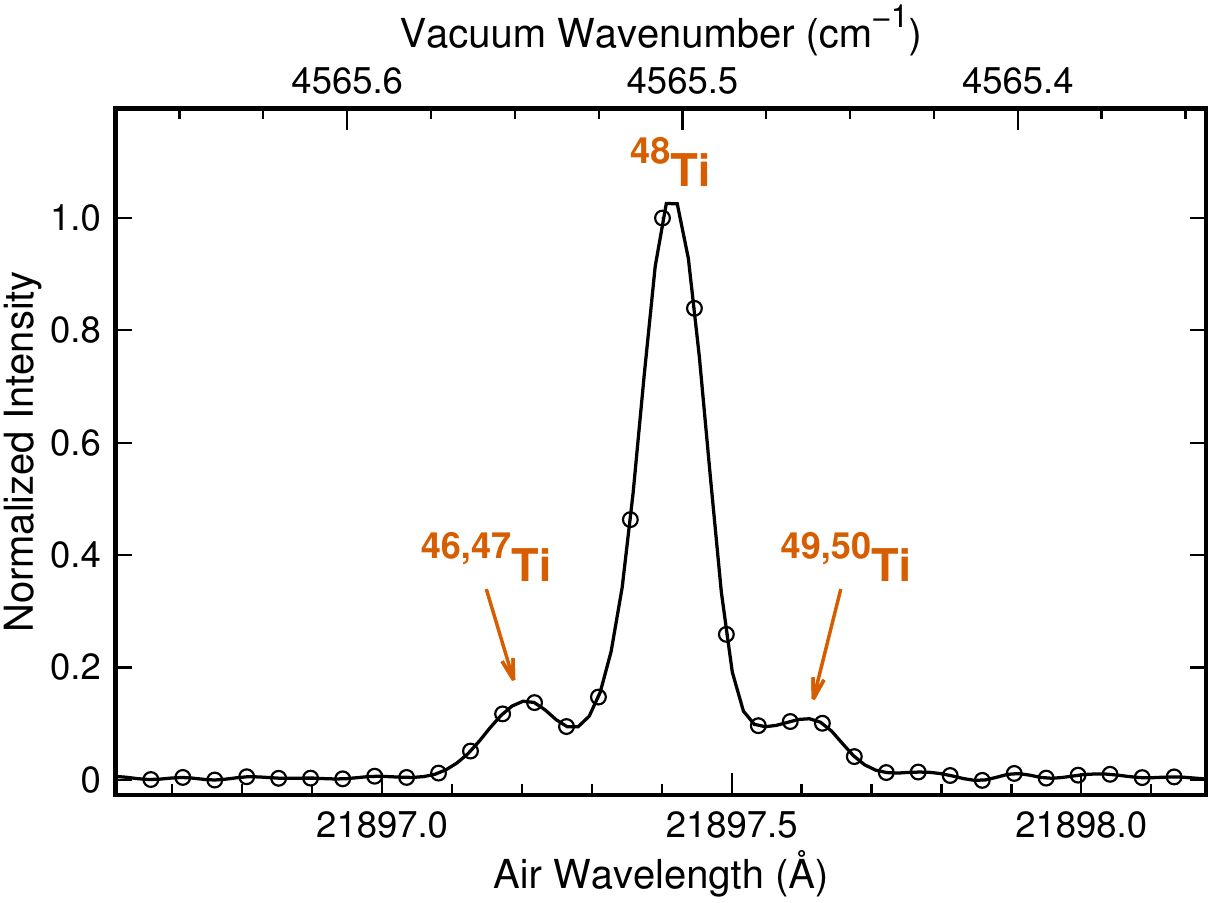}
\caption{
\label{Ti_iso}
\footnotesize
Example laboratory spectrum for one of the Ti I IR lines included in this study.  
The main component of the line is the dominant $^{48}$Ti isotope, while the side peaks 
correspond to the minor Ti isotopes.  This spectrum was recorded using the 
NIST 2-m FTS and a high-current hollow cathode lamp, and is representative 
of the type used to derive our Ti I isotope shifts.
 }
\end{figure}

We recorded the \species{Ti}{i} ISs in the laboratory using the NIST 2-m 
Fourier Transform Spectrometer (FTS) and a high-current hollow cathode lamp.  
Although our laboratory studies are limited to data from FTS,
there have been laboratory studies of ISs in \species{Ti}{i} employing 
narrow band lasers (\citealt{azar92}, \citealt{anas94}, \citealt{luc94}, 
\citealt{gang95} \citealt{luc96}, \citealt{furmann96}, \citealt{jin09}).  
Single frequency laser experiments can fully resolve HFS structure of the 
odd isotopes \iso{47}{Ti} and \iso{49}{Ti}. 
Small FSs in \species{Ti}{i} have been studied (\citealt{azar92}, 
\citealt{anas94}, \citealt{luc94}). 
A J-dependence of ISs for lines in a single multiplet have been detected 
and studied (\citealt{gang95}). 
Such J-dependences are typically quite small changes in ISs that are already 
small, but do occur due to mixing of levels. 
Negative ISs, in which the heavier isotopes are the red of the lighter 
isotopes, were clearly observed and explained by \cite{luc96}. 
The dominance of the SMSs in the ISs of 
3$d^{3}$ 4$s$ $a$$^{5}F$ $-$ 3$d^{2}$ 4$s$4$p$ $y$$^{3}F$
is responsible for negative ISs as demonstrated using Hartree-Fock calculations 
by \citeauthor{luc96}. 
Of our 10 \species{Ti}{i} IGRINS lines, the six longest wavelength transitions
belong to the 3$d$$^{3}$ 4$s$ $a$$^{5}$$P$ $-$ 3$d$$^{2}$ 4$s$4$p$ $z$$^{5}$$D$ 
multiplet and exhibit the negative ISs explained by \citeauthor{luc96} 
in terms of the SMSs of 3$d$ $-$ 4$p$ transition. 
In our FTS data we find the \iso{49,50}{Ti} isotopes of these lines at $-$0.0430 
$\pm$ 0.0108 cm$^{\rm -1}$ with respect to the dominant \iso{48}{Ti} isotope 
and the \iso{46,47}{Ti} isotopes at $+$0.0430 $\pm$ 0.0132 cm$^{\rm -1}$ 
with respect to the dominant \iso{48}{Ti} isotope. 
We are not able to detect any J-dependence in our FTS data on this IR multiplet.
For all of its transitions, in our laboratory spectra
we are unable to resolve the minor odd isotopes \iso{47}{Ti} and \iso{49}{Ti}
from their even isotope companions \iso{46}{Ti} and \iso{50}{Ti}, 
respectively.
Since we could not definitively separate the contributions to the 
shifts from the even isotopes, we quote the combined shift due to 
both isotopes.
An example laboratory spectrum for one of the \species{Ti}{i} IR lines is 
presented in Figure~\ref{Ti_iso}.
In solar-system material the four minor isotopes are about equal abundance, 
and are displaced nearly symmetrically to the blue and red of \iso{48}{Ti} 
as seen in this figure, and the wavelength shifts are fractions 
of the total line widths in stellar spectra.  
Therefore an isotopic mix very different from the solar system would be 
needed in another star to detect noticeable line asymmetries.
 
The next two shorter wavelength transitions in our IGRINS data, at 17376 \AA\ 
and 17383 \AA, connect rather high lying levels of neutral Ti and do not have 
sufficient S/N in our FTS data for extracting ISs.
  
The two shortest wavelength \species{Ti}{i} transitions 
in our IGRINS data, at 15543 \AA\ and 15602 \AA, are the  
3$d$$^{3}$ 4$s$ $a$$^{3}$$G$$_{4}$ $-$ 3$d$$^{2}$ 4$s$4$p$ $z$$^{3}$$G$$_{4}$ 
and
3$d$$^{3}$ 4$s$ $b$$^{1}$$G$$_{4}$ $-$ 3$d$$^{2}$ 4$s$4$p$ $z$$^{1}$$G$$_{4}$ 
lines.
These lines also have negative ISs but they are not as well resolved 
in our FTS data. 
This illustrates the smaller fractional ISs as the optical wavelength region 
is approached. 
We find the \iso{49,50}{Ti} isotope shift for the 15543 \AA\ line to be
$-$0.0396 $\pm$ 0.0050 cm$^{\rm -1}$ with respect to the dominant \iso{48}{Ti} 
isotope and the \iso{46,47}{Ti} isotope shift to be
$+$0.0408 $\pm$ 0.0050 cm$^{\rm -1}$
with respect to the dominant \iso{48}{Ti}\ isotope.
For the 15602 \AA\ line, the \iso{49,50}{Ti} shift with respect to \iso{48}{Ti} 
is $-$0.0356 $\pm$ 0.0056 cm$^{\rm -1}$,
and the \iso{46,47}{Ti} shift with respect to \iso{48}{Ti} 
is $+$0.0361 $\pm$ 0.0092 cm$^{\rm -1}$.

Embedding this \species{Ti}{i} isotopic structure into the atomic line list 
cures the broadening observed at the wings of these weak lines in the solar 
spectrum, providing a more robust test bed for the oscillator strengths 
reported by \cite{blackwell06}. 
We applied the \loggf\ values of \citeauthor{blackwell06} to the solar 
spectrum for the seven lines in common (except for the Ti I line at 
15602.84~\AA) and remeasured the solar Ti abundance.
These computations indicated that the abundances from two of the lines 
at 15543.76 and 21782.94 \AA\ result in a considerable amount of offset 
from the Ti abundance reported by \cite{asplund09} 
(\logeps(Ti)$_{\odot}$ = 4.95), with the latter suggesting
about 0.25 dex lower Ti abundance in the Sun.  
Therefore, to achieve internally consistent Ti abundances in the Sun, Arcturus and our 
stars we opted to apply reverse solar analysis to all \species{Ti}{i} lines 
we list in Table~\ref{tab-lines}. 
Our astrophysical \loggf\ values stay nearly within the \loggf\ 
uncertainties given by \citeauthor{blackwell06}, except for the two lines mentioned here.

\subsection{$\alpha$ and Odd-Z elements}\label{alphaodd-ir}

Molecular lines become major contributors in the spectrum of cool stars
towards longer wavelengths. 
As a result, the near-$IR$ spectra contains many molecular lines,
mainly CN, OH and CO. 
Therefore, applying the SYN technique is advisable
for all $IR$ features of this study.

We derived the abundances of $\alpha$ elements 
Mg, Si, Ca and S from their neutral species transitions.
We plot example Mg, Ca and S lines in \ffor\ in Figure~\ref{hip54CaSMg}.
The line-by-line $\alpha$ abundances for \ffor\ are 
displayed in Figure~\ref{alphas}. 
We also include in this figure $\alpha$-like Ti, for which we have both
neutral and ionized species transitions.
By inspection it is clear that the abundances from lines in the $H$ and $K$
bands are in accord with their optical counterparts.
This conclusion is confirmed by the abundance statistics of
Tables~\ref{tab-54048}$-$\ref{tab-114809}.
The $IR$ abundances often have better internal self-consistency 
compared to optical ones. 
Additionally, for some elements the $IR$ abundances may be preferrable to
the optical ones.
For example, we were able to use 11 \species{Mg}{i} lines in the infrared but 
only two could be used in the optical.
The 5528 and 5711~\AA\ lines are both very strong in our RHB stars.
For \ffor, their reduced equivalent width RW ($\equiv$~log(EW/$\lambda$))
values are $-$4.4 and $-$4.7, respectively. 
This means that both are well up on the flat part of the curve-of-growth, 
relatively insensitive to Mg abundances.
However, the $IR$ \species{Mg}{i} lines include several on the linear part 
of the curve-of-growth; we trust the $IR$ Mg results.

We report S abundances for our program stars for the first time.  
Considering the importance of this element for the 
cosmochemical evolution of the Galaxy, it is important to compare the
abundances derived from \species{S}{i} lines in the $IR$ to their very 
few optical counterparts. 
There are 10 useful \species{S}{i} lines available in $H$ and $K$ bands
for our RHB stars (Table~\ref{tab-lines}).

The four true $\alpha$ elements have virtually the same abundance levels.
Defining $\langle$[$\alpha$/Fe]$\rangle$ $\equiv$ $\langle$[Mg,Si,S,Ca/Fe]$\rangle$,
we find that $\langle$[$\alpha$/Fe]$\rangle$~$\simeq$0.15 with 
$\sigma$~$\simeq$~0.08 in all three RHB stars in both optical and $IR$ spectral
regions.

The abundances of odd-Z elements Na and Al were also obtained from
the near-$IR$ lines (Figure~\ref{other_elem}). 
Additionally, we were able to measure the abundances
of another two odd-Z elements, P and K,
for the first time for our targets.

Na abundances were determined from \species{Na}{i} lines at 22056.4, 
22083.7, 23348.4 and 23379.1 \AA. 
The \loggf\ values of Na lines are quite well 
determined as listed in the NIST database (ranked A+). 
The transitions at 23348.4 and 23379.1 \AA\, on average, 
give about 0.1 dex higher abundances in all RHB stars (Table~\ref{tab-lines}, 
Figure~\ref{other_elem}) than do other \species{Na}{i} lines.
The 23348.4 \AA\ line is also partly blended with a CO line.
Additionally, the effects of nLTE on these lines are yet to be investigated. 
The Na abundances obtained from the $IR$ are in agreement
with the optical Na abundances and show the same trend in all RHB stars.
Na abundances are well enhanced in \ffor\ and \oneone\ by 
[Na/Fe]~$>$~0.25 dex. \fsev, on the other hand, has solar Na abundance.

Al abundances were measured from six \species{Al}{i} lines; two in the $H$,
and four in the $K$ band. 
When compared to other elements investigated in 
this section, the Al abundances indicate a higher line-to-line scatter 
($\sigma$~$\simeq$~0.1 dex),
and the $K$ band Al lines are the main contributors to this scatter. 
$H$ band Al abundances from 16763.4 and 17699.0 \AA\
lines are usually in agreement with the optical results. 
The average infrared Al abundance is always $\sim$0.1 dex 
higher compared to the average optical Al abundance in case of all RHB stars, 
with slight overabundances in \ffor\ and \oneone.

The most commonly used P lines for abundance calculations are the 
\species{P}{i} $UV$ doublet, 2135/2136~\AA\ (e.g. 
\citealt{jacobson14,roederer14}) and two near $IR$
\species{P}{i} lines at 1050$-$1082 nm (e.g. \citealt{caffau11,maas17}).
In this study, we report the abundances from two different near-$IR$ 
lines located in the $H$ band: 15711.5 and 16482.9 \AA\ (Figure~\ref{PK}). 
We could not derive an abundance from the 16482.9~\AA\ 
line for \oneone. 
It is too weak in our IGRINS spectrum, and only an upper limit 
could be derived from the \species{P}{i} line at 15711.5 \AA.
P abundances are estimated to be at solar values in all RHB stars. 

$H$ band data provide two \species{K}{i} lines at 
15163.1 and 15168.4 \AA\ (Figure~\ref{PK}). 
This region, however, is fairly contaminated by the CN molecular lines
that can be seen in the figure.
Nevertheless, they provide K abundances around solar as expected 
from mildly metal-poor and solar metallicity stars 
(e.g. \citealt{takeda02,takeda09}). 
As discussed in \S\ref{alphaodd-opt}, nLTE effects on the optical lines are 
severe and the K abundances obtained from those lines 
are about 0.6 dex higher than the solar value. Considering the K abundances we 
determined from the $H$ band lines, we suggest that 15163.1 and 
15168.4~\AA\ \species{K}{i} lines may provide less nLTE-affected abundances.

\subsection{Fe-group and neutron-capture elements}\label{fencap-ir}

Abundances of the same Fe-group elements discussed in \S\ref{fencap-opt} 
were also determined from the $IR$ lines of Sc, Ti, Cr, Co and Ni 
(Figure~\ref{other_elem}). 
Here we also give the Fe abundances from 
the \species{Fe}{i} lines identified in the $H$ and $K$ bands.

We could identify only two \species{Sc}{i} lines in the $K$ band: 
22065.2 and 22266.7 \AA.
Recently, \cite{pehlivan15} measured the oscillator strengths for single atomic 
level transitions of \species{Sc}{i} lines and \cite{vandeelen17} studied the 
HFS structure of these lines in the near-$IR$. We adopted the oscillator 
strengths and HFS splittings from these studies and applied SYN 
analysis to our Sc lines. 
These lines are weak in our stars and we could only detect them in \ffor. 
Optical abundances were determined from \species{Sc}{ii} lines. 
The results for \ffor\ from both infrared and optical species have a small 
offset: 
[\species{Sc}{ii}/Fe]$_{\rm opt}$$-$[\species{Sc}{i}/Fe]$_{\rm ir}$ = 0.15.
This discrepancy somewhat resembles the difference observed between the 
neutral and ionized species of Cr in the optical.

We identified 10 useful \species{Ti}{i} lines and one 
\species{Ti}{ii} line at 15783~\AA\ \citep{wood14c} in our IGRINS data.
The abundances we obtained from the infrared are in accord with the 
ones obtained from the optical. 
Both the optical and $IR$ Ti lines in \fsev\ support a modest abundance
deficiency: [Ti/Fe]~$\simeq$~$-$0.15.
\ffor\ and \oneone\ have about solar Ti abundances.
Although the general [\species{Ti}{ii}/Fe] abundance trend 
is the same both in the infrared and optical, there is a mean offset of about 
$+$0.16 dex between the abundances [\species{Ti}{ii}/Fe]$_{\rm ir}$ and 
$\langle$[\species{Ti}{ii}/Fe]$\rangle$$_{\rm opt}$ in all RHB stars. 

The other Fe-group elements have general agreement
between optical and $IR$ transitions.
\species{Cr}{i} lines and the abundances are about solar in all RHB stars.   
Note that there is just one \species{Co}{i} line identified at 16757.6 \AA.
Six \species{Ni}{i} lines were identified in the $H$ band. The Ni abundances 
for all RHB stars are in good accord with the optical ones.

There are many \species{Fe}{i} lines 
in the both $H$ and $K$ bands (Figure~\ref{atlas}). 
Here we only discuss the ones that are not heavily blended with other
atomic and molecular features.
In total, we were able to identify 27 \species{Fe}{i} lines in both bands. 
We applied both EW and SYN methods to measure the Fe abundances in our 
stars (please see  \S\ref{fei} for further discussion on the EW method). 
In Tables~\ref{tab-lines}$-$\ref{tab-114809} we only list
the abundances obtained from the SYN method.
In Table~\ref{tab-synew} we compare the Fe abundances derived from the 
same \species{Fe}{i} lines applying both methods.
The agreement between the results from two methods is well within the 
uncertainty of 0.05 dex. 
Comparison of both optical: \logeps(\species{Fe}{i})$_{\rm opt}$ = 7.17 
(Af{\c s}ar18a) and infrared: 
\logeps(\species{Fe}{i})$_{\rm ir(SYN)}$ = 7.18 abundances
results in a very good agreement.
In \S\ref{fei} we discuss other aspects of the $IR$ 
\species{Fe}{i} lines.

We derived the abundances of \ncap\ elements Ce, Nd and Yb from their
ionized species transitions. 
We adopted the \loggf\ values of \species{Ce}{ii} from \cite{cunha17} 
and the \loggf\ value for \species{Nd}{ii} was taken from \cite{hassel16}.
Ce and Nd have been already discussed in 
\S\ref{fencap-opt}, but Yb was detected only in the $H$ band at 
16498.4 \AA.
The commonly used near-$UV$ \species{Yb}{ii} resonance line at 3694 \AA\ 
was unavailable for our RHB stars because it is too strong and lies in a very 
crowded, low-flux spectral region.
In Figure~\ref{ncapts}, we display one \species{Ce}{ii} line and the 
\species{Yb}{ii} line in our stars. 
Yb is a \ncap\ element whose solar system abundance mainly comes
from rapid \ncap\ (\rpro) nucleosynthesis. 
\cite{hawkins16} detected the \species{Yb}{ii} line in the Arcturus spectrum.
As is seen in their Figure 13, the Yb line in Arcturus is considerably blended
with a CO feature. However, this contamination is fairly weak in our stars due to
their relatively high temperatures compared to Arcturus. Nonetheless, one needs
to be aware of the potential CO contribution to the Yb line in cooler stars.
Yb is essentially solar in all of our RHB stars, 
$\langle$[Yb/Fe]$\rangle$~$\simeq$~$+$0.06, while the \spro\ elements
Ce and Nd exhibit large star-to-star scatter.
This issue will be investigated further in a future paper with a larger
RHB sample.

\subsection{The CNO Group}\label{cno-ir}

The $H$ and $K$ bands contain molecular bands
of OH, CN  and CO, which we used to determine the abundances of C, N and O.
Since the CNO elements are tied together through molecular equilibrium, 
we applied an iterative SYN analysis for the CNO abundances
as described in Af{\c s}ar16. 
Since O is expected to be more abundant than C it should be less affected
by CO formation, so we started with OH analyses to determine the O abundances. 
However, OH lines are very weak in these warm RHB stars.
We were able to detect up to five useful OH transitions in our spectra. 
The OH region at $\sim$16872 \AA\ is given in Figure~\ref{OH} for all RHB stars
and Arcturus. The high-resolution Arcturus atlas (\citealt{hinkle95}) shows
that there are in fact two distinct molecular OH features centered at 
$\sim$16872 \AA. 
The same features appear as blends in our stars.
\ffor\ has the lowest temperature (\teff~=~5099~K) of our stars and the 
OH feature at 16872 \AA\ is better resolved compared to the one 
in \fsev\ (\teff~=~5307~K) at the same metallicity.
Although \oneone\ spectrum has lower $S/N$ and the lowest metallicity 
([M/H]~=~$-$0.38) in our sample, its OH feature still is present.
Abundances of O derived from the OH features in all three stars agree well with 
those obtained from the [\species{O}{i}] lines in the optical:
$\langle$[O/Fe]$_{\rm opt}$$-$[O/Fe]$_{\rm ir}$$\rangle$ = 0.04 dex
(Table~\ref{tab-54048}$-$\ref{tab-114809}).

C abundances were determined from the numerous CO 
molecular lines in the $K$ band (Figure~\ref{kspec}). 
We started the abundance analysis by adopting the O abundance 
derived from OH lines. 
The C abundances obtained from the CO lines came out to be about 0.1 dex 
higher than the C abundances estimated from the CH lines and lines of the 
C$_{2}$ Swan band heads discussed in \S\ref{light-opt}. 
This very small difference may be attributed to the larger 
temperature sensitivity of CO molecular line strengths (see Af{\c s}ar16 for 
a detailed discussion). 
On the other hand, CH and C$_{2}$ molecular lines come with their own problems.
They are very small constituents in the overall
C molecular equilibrium, and they are known to be severely blended with 
atomic absorption features; all these factors bring difficulties in 
determination of C abundances.
Considering that we can only work with the molecular
features to determine C abundances in both optical and $IR$ regions,
the agreement in abundances obtained from both regions is reasonable.

In addition to the high-excitation potential \species{C}{i} lines investigated 
in \S\ref{light-opt}, we located four useful \species{C}{i} lines at 
16021.7, 16890.4, 17456.0, and 21023.2 \AA\ in the near-$IR$. 
These lines are the ones least blended
with surrounding features. 
First we collected the \loggf\ values from the NIST database and applied 
them to the solar spectrum using the SYN method.
Except for the \species{C}{i} line at 21023.2 \AA, 
other \loggf\ values provided solar C abundances about $-$0.11 dex lower
than the abundance reported by \cite{asplund09}. 
Applying NIST \loggf\ values to our RHB stars and Arcturus, however, 
resulted in very good agreement with the C abundances obtained from CO features. 
The average C abundances from \species{C}{i} lines are,
$\langle$\logeps(C)$\rangle$$_{\rm NIST}$ = 7.72, 8.16, and 7.84 
for \ffor, \fsev, and \oneone, respectively. 
To ensure the internal consistency of this study, we decided
to apply reverse solar analysis to the \species{C}{i} lines and obtained 
astrophysical \loggf\ values for the lines at 16021.7, 16890.4, and 17456.0~\AA.
Applying astrophysical \loggf\ values to our stars and Arcturus yielded 
C abundances, surprisingly, still in good agreement with the ones obtained 
from CO features. 
The average C abundances determined using 
astrophysical \loggf\ values for our stars are
$\langle$\logeps(C)$\rangle$$_{\rm astro}$ = 7.78, 8.21, and 7.90 
for \ffor, \fsev, and \oneone, respectively. 
The \species{C}{i} line at 8335.1 \AA\ in the optical 
and the $IR$ lines at 16890.4 and 17455.9 \AA\ increase the abundance 
scatter obtained from \species{C}{i} lines. 
Resolution of this issue is beyond the scope of our study.
In a followup study we will
perform a detailed investigation with a larger sample for better 
clarifying the behavior of the $IR$ \species{C}{i} lines. 
Overall results for the C abundances from all 
individual species and the average values from the five species abundance
indicators for the stars are listed in Tables~\ref{tab-c}
and \ref{tab-cmean}.

We derived the N abundances from the CN molecular transitions in the $H$ band. 
The C and O abundances derived from CO and OH lines 
were taken from the analyses described above.
$H$ band CN features provide relatively consistent 
N abundances with the ones obtained from the optical CN features.

Prominent features including band heads of the ground 
electronic state first overtone (2$-$0) and (3$-$1) R-branch vibration-rotation bands of $^{12}$CO (2$-$0) 
are accompanied by $^{13}$CO (2$-$0) features near 23440 \AA\ and those of
$^{13}$CO (3$-$1) at about 23730 \AA.  
These CO bands offer the opportunity to derive more accurate \carbiso\ ratios
than possible with the weak $^{13}$CN bands in the optical region,
as discussed in \S\ref{light-opt}.  
This is illustrated in Figure~\ref{carbiso} with side-by-side comparisons
of CN (left panel) and CO (right panel) features.
In Table~\ref{tab-carbiso} we list the derived \carbiso\ values for the
three RHB stars.
Happily, the ratios derived from both molecules in fairly good agreement.
Most importantly, the $^{13}$CO band head features let us prove that 
the \oneone\ surface has been enriched with CN-cycle products 
($\langle$\carbiso$\rangle$ = 15) which was not possible in ASF12.
This improvement will allow one to redefine and constrain the \carbiso\ ratios 
in a larger number of evolved stars, which in turn will bring more constrains 
on stellar evolution models.

\section{\species{Fe}{i} LINES IN THE INFRARED AS TEMPERATURE INDICATORS}\label{fei}

To derive the model atmospheric parameters without having any information 
from the optical region is a difficult task. 
The most important starting point is to have an accurate value
of effective temperature.
Here, using our limited sample, we describe a couple of methods to estimate  
effective temperatures of stars using only the $IR$ spectral features. 
The second paper of this study will also explore the methods to derive 
model atmospheric parameters using only the spectral features in the $IR$ 
with a larger sample.

As mentioned earlier, spectral regions redward of
about 7000~\AA\ have substantial molecular features in G-K giant stars;
truly unblended atomic absorption lines become increasingly difficult to find
as wavelength increases.
Nonetheless we searched for as many clean \species{Fe}{i} lines as possible
in our IGRINS spectra, using the atomic data sources cited above and 
the Arcturus line list of \cite{hinkle95}.
We were able to identify about 25 relatively unblended
\species{Fe}{i} lines in total. 
We first measured the EW values of these lines in Arcturus and derived
individual line abundances, again adopting the atmospheric parameters 
from \cite{ramirez11}. 
As seen in the left panel of Figure~\ref{FeTeff} the derived Fe
abundances show no obvious trends with excitation potential $\chi$ and 
with RW.
The average abundance obtained from the EW measurements of infrared \species{Fe}{i} lines are 
in accord with the Fe abundance reported by \citeauthor{ramirez11}:
\logeps(Fe)$_{\rm EW}$ = 6.96 ($\sigma$ = 0.02), and \logeps(Fe)$_{\rm Ramirez}$ = 6.98.
We then applied the same procedure to our program stars. 
As an example, Fe line abundances for \oneone\ are shown in the right panel of 
Figure~\ref{FeTeff}. 
No obvious trends with $\chi$ and RW are seen. 
The average \logeps(Fe)$_{\rm ir}$ values for our RHB stars (Table~\ref{tab-synew})  
are in good agreement with the \logeps(Fe)$_{\rm opt}$.
These efforts suggest that the infrared \species{Fe}{i} lines can be used 
to determine reliable stellar effective temperatures in the cases of no 
available data from the optical region.

We also applied \citeauthor{fukue15} LDR relations to our stars and obtained 
\teff$_{(\rm LDR)}$ values from eight pairs of absorption lines. We excluded 
the \species{K}{i}(15163.09)/\species{Fe}{i}(15244.97) pair in all three cases 
because it always yielded about 400 K higher temperature values compared 
to other absorption line pairs. The \teff$_{(\rm LDR)}$ values we obtained 
by applying \citeauthor{fukue15} method for our stars:
\ffor, \teff$_{(\rm LDR)}$ = 5157 K ($\sigma$ = 123 K);
\fsev, \teff$_{(\rm LDR)}$ = 5312 K ($\sigma$ = 215 K);
\oneone, \teff$_{(\rm LDR)}$ = 5235 K ($\sigma$ = 151 K).
These are in excellent agreement with our optical spectroscopic \teff\
values, being on average only about 50~K higher, well within the mutual
uncertainties of both sets of measurements.

\section{SUMMARY AND DISCUSSION}\label{disc}

This paper is the first one of a series of papers that explores stellar 
atomic and molecular features of Galactic field RHB and red giant
stars observed in the near-$IR$ region. 
IGRINS high-resolution, high S/N $H$ and $K$ band spectra bring 
valuable new information for this purpose.
Several important light elements are poorly studied 
(\eg, S and K) or not available (\eg, P) in high-resolution optical spectra.
IGRINS near-$IR$ spectra provide more robust abundance analysis for 
many elements.

We have conducted, for the first time, the abundance analysis of three 
RHB stars; \ffor, \fsev\ and \oneone, 
using high-resolution, high S/N near-$IR$ IGRINS spectra. 
Detailed abundance analyses were carried out for 21 species of 20 elements 
including $\alpha$ (Mg, Si, S and Ca); odd-Z (Na, Al, P and K); 
Fe-group (Sc, Ti, Cr, Co, Fe, and Ni); $n$-capture elements (Ce, Nd, Yb); 
and CNO group elements. 
We also used the ground electronic state first overtone ($\Delta$$v$ = 2) 
R-branch vibration-rotation bands of (2$-$0) and (3$-$1), which are the prominent features in K-band 
spectra, to determine \carbiso\ ratios from $^{13}$CO (2$-$0) features near 23440 \AA\ and 
$^{13}$CO (3$-$1) features at about 23730 \AA.

To perform the abundance analysis in the $H$ and $K$ band spectra,
we adopted the model atmosphere parameters from Af{\c s}ar18a.
Using the high-resolution optical spectra of our RHB stars (ASF12), we also
determined the abundances of the same elements (except for P) in the optical 
region. This broad wavelength coverage allowed us too make more comprehensive 
analyses of our stars.

\begin{figure}
\epsscale{0.85}
\plotone{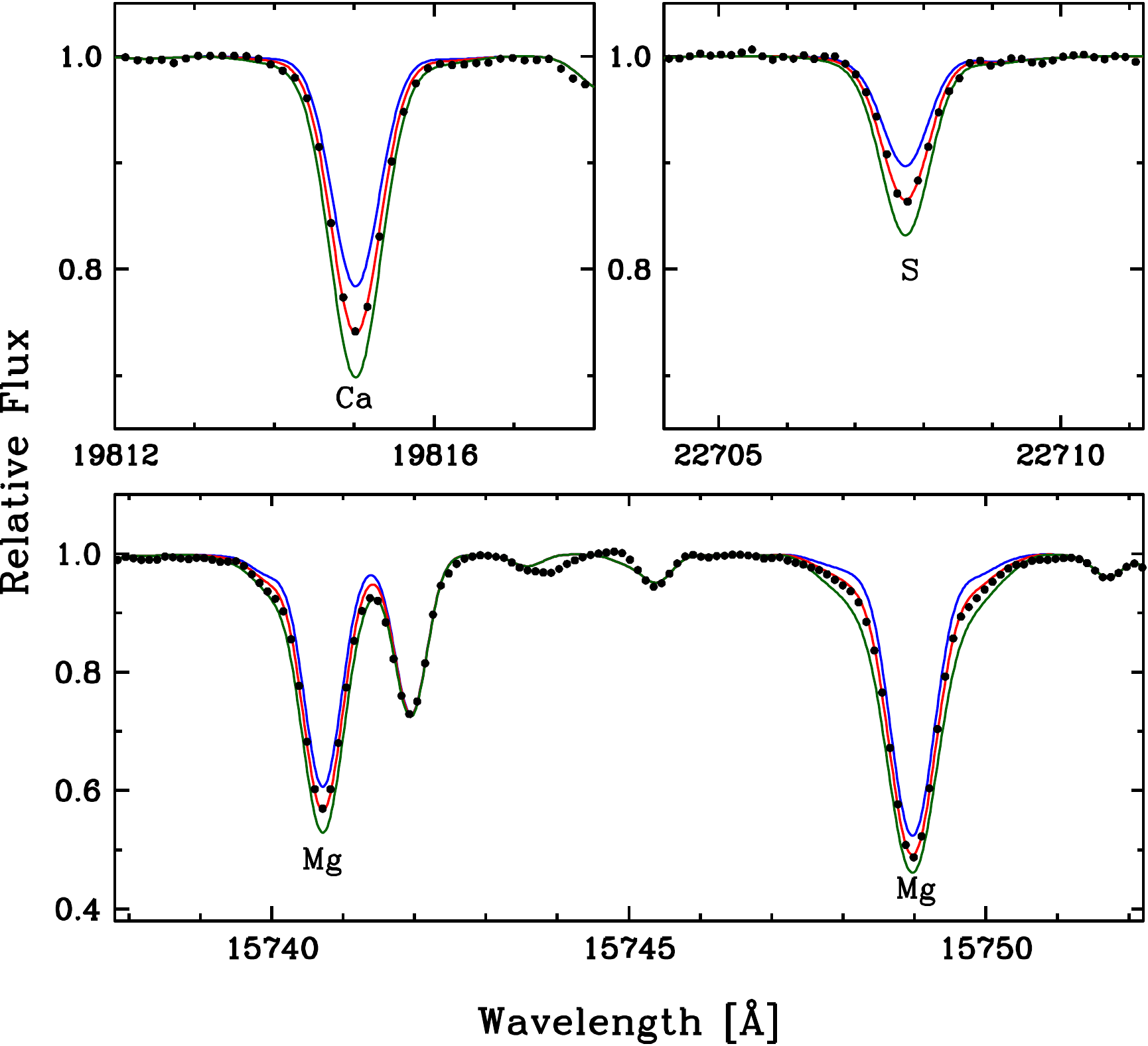}
\caption{
\label{hip54CaSMg}
\footnotesize
Example regions for Mg, Ca and S in \ffor. The syntheses represented with 
blue and green lines depart from the best fit (red line) by $\pm$0.3 dex. 
The lines, colors and symbols are the same as in Figure~\ref{carbiso}.
}
\end{figure}

\begin{figure}
\epsscale{0.75}
\plotone{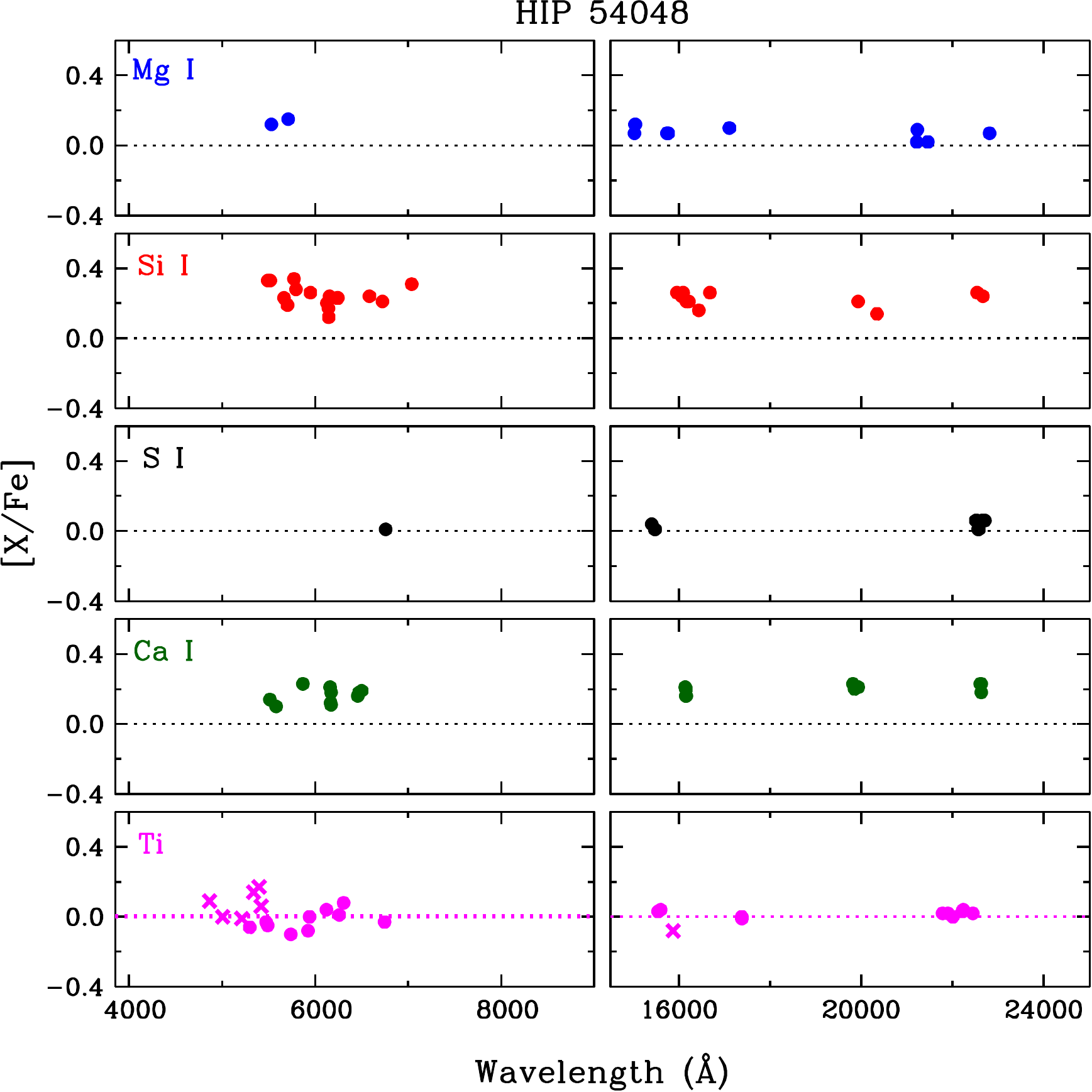}
\caption{
\label{alphas}
\footnotesize
Relative abundance ratios [X/Fe] of $\alpha$ elements in \ffor.
Ti panel includes both \species{Ti}{i} (dots) and \species{Ti}{ii} (crosses).
We included Ti in this plot since it is sometimes regarded as an 
$\alpha$ element.
}
\end{figure}

The abundances of $\alpha$ elements obtained from both wavelength 
regions are generally in good agreement and, on average, they 
indicate $\alpha$ enhancements of about 
$\langle$[$\alpha$/Fe]$\rangle$~$\simeq$~0.15 dex. 
Our IGRINS spectra yield substantial numbers
of Mg and S lines with internally self-consistent abundance 
results (Figure~\ref{alphas}).
We have identified 11 \species{Mg}{i} lines in contrast to two optical 
\species{Mg}{i} lines that come with analytical difficulties due to their 
large absorption strengths.
The $IR$ \species{Mg}{i} lines now give us more robust Mg 
abundances in these stars. 
Being located closer to the linear part of the curve-of-growth than
the two optical lines, we trust the abundances of Mg obtained from the 
$IR$ \species{Mg}{i} lines (Figure~\ref{hip54CaSMg}).

Sulfur is one of the fundamental building block elements 
of life in the universe along with C, H, N, O and P. 
Sulfur and sulfur-containing compounds are common constituents in 
Solar System material. 
High amounts of sulfur are produced in supernova nucleosynthesis during 
explosive oxygen burning (\citealt{truran73}). Sulfur is volatile with its low 
condensation temperature; it can not be condensed into interstellar dust grains.
Therefore, its depletion-free nature makes sulfur a valuable tracer of 
Galactic nucleosynthesis; understanding of its 
contribution to Galactic chemical evolution is important. 
There have been a number of studies that investigate the behavior of S in 
the Galaxy as one of the $\alpha$ elements.  
Different investigations claim different results. 
Some suggest that S behaves
like other $\alpha$ elements, showing a [S/Fe] \textit{plateau} at certain 
metallicities, while others indicate that [S/Fe] has a linear increase 
with decreasing metallicities (e.g. \citealt{caffau05}, and references therein).
There are few well-known \species{S}{i} lines in the optical and shorter 
wavelengths of near-$IR$: 6743$-$6757 \AA\ optical multiplet, 
8694$-$5 \AA, 9212$-$9237 \AA\ $IR$ triplet, and 10455$-$9 \AA\ $IR$ triplet. 
Most of these lines were investigated by \cite{korotin09} and \cite{takeda16} 
for their being subjected to nLTE effects. 
All these S lines appear to need nLTE corrections in different amounts,
6743$-$6757 \AA\ optical multiplet having the smallest (to $-$~0.10 dex) 
nLTE correction among others.

IGRINS provides more \species{S}{i} lines beyond those
of the optical and very near-$IR$. 
We were able to identify 10 useful $H$ and $K$ band \species{S}{i} lines.
We also measured the S abundance using the optical triplet centered at 
6757.0 \AA. 
The internal abundance self-consistency of all \species{S}{i} lines 
is encouraging.
The $IR$ S lines are mostly weak and likely to be less affected by nLTE 
conditions.
These lines, studied here for the first time in these stars, promise
to provide more robust investigation at a range of metallicities. 
In the forthcoming paper of this series, we will investigate S
in more detail with a larger sample.

The abundances of odd-Z elements Na, Al, P and K 
were computed from both optical (when available) and infrared regions. 
Optical transitions of these elements are subject to substantial nLTE
corrections (e.g. \citealt{takeda02,takeda03,lind11,nord17,mucciarelli17}). 
Among these, optical \species{Al}{i} lines used in this study are the 
least affected ones by nLTE, while \species{K}{i} at 7699 \AA\ is the 
most affected, by up to $-$0.7 dex (\citealt{takeda02,takeda09}).
This is also found in our stars (Tables~\ref{tab-54048}$-$\ref{tab-114809}).
 
The Na abundances obtained in both wavelength regions are in good agreement 
in all three RHB stars but the star-to-star scatter is substantial. 
Even with nLTE corrections to the optical \species{Na}{i} 
lines, overabundances observed in both \ffor\ and \oneone\ remain. 
The good agreement between the optical and $IR$ Na abundances suggests 
that all of these lines might be affected by nLTE conditions about the 
same amount.  
A similar situation is also observed in Al abundances from both regions. 
\ffor\ and \oneone\ also show slight Al overabundances, reminiscent of
the well known [Na/Fe]-[Al/Fe] correlation observed in globular clusters 
(e.g. \citealt{kraft94} and references therein).

\begin{figure}
\epsscale{0.85}
\plotone{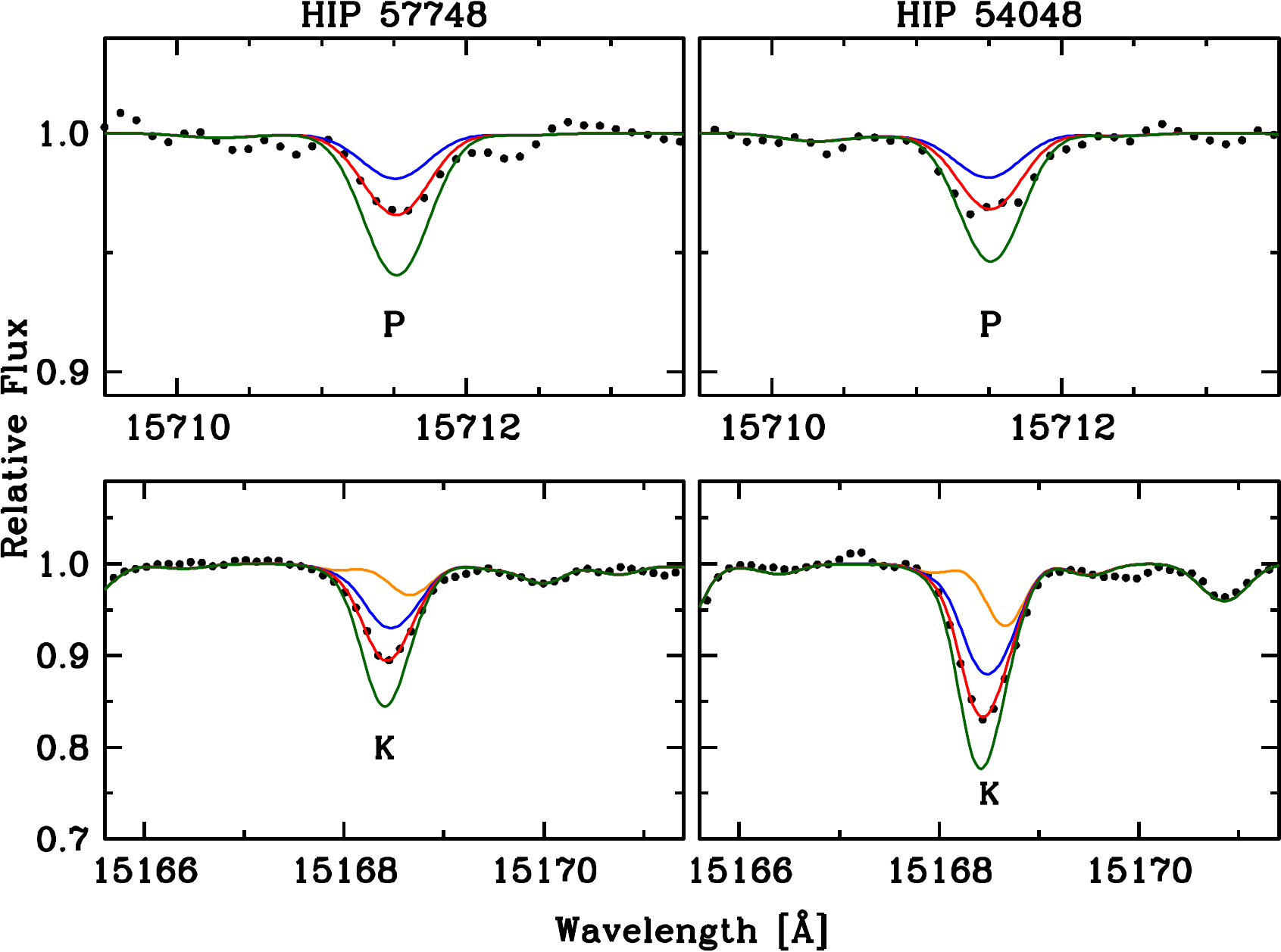}
\caption{
\label{PK}
\footnotesize
One of the P and K regions in \ffor\ and \fsev\ illustrated as an example. 
The syntheses illustrated with blue and green lines depart 
from the best fit (red line) by $\pm$0.3 dex. 
K lines are mainly blend with CN features. Orange lines assume
``no K detection'' to draw attention to the effect of the blended feature.
The lines, colors and symbols are the same as in Figure~\ref{carbiso}.
}
\end{figure}

\begin{figure}
\epsscale{0.75}
\plotone{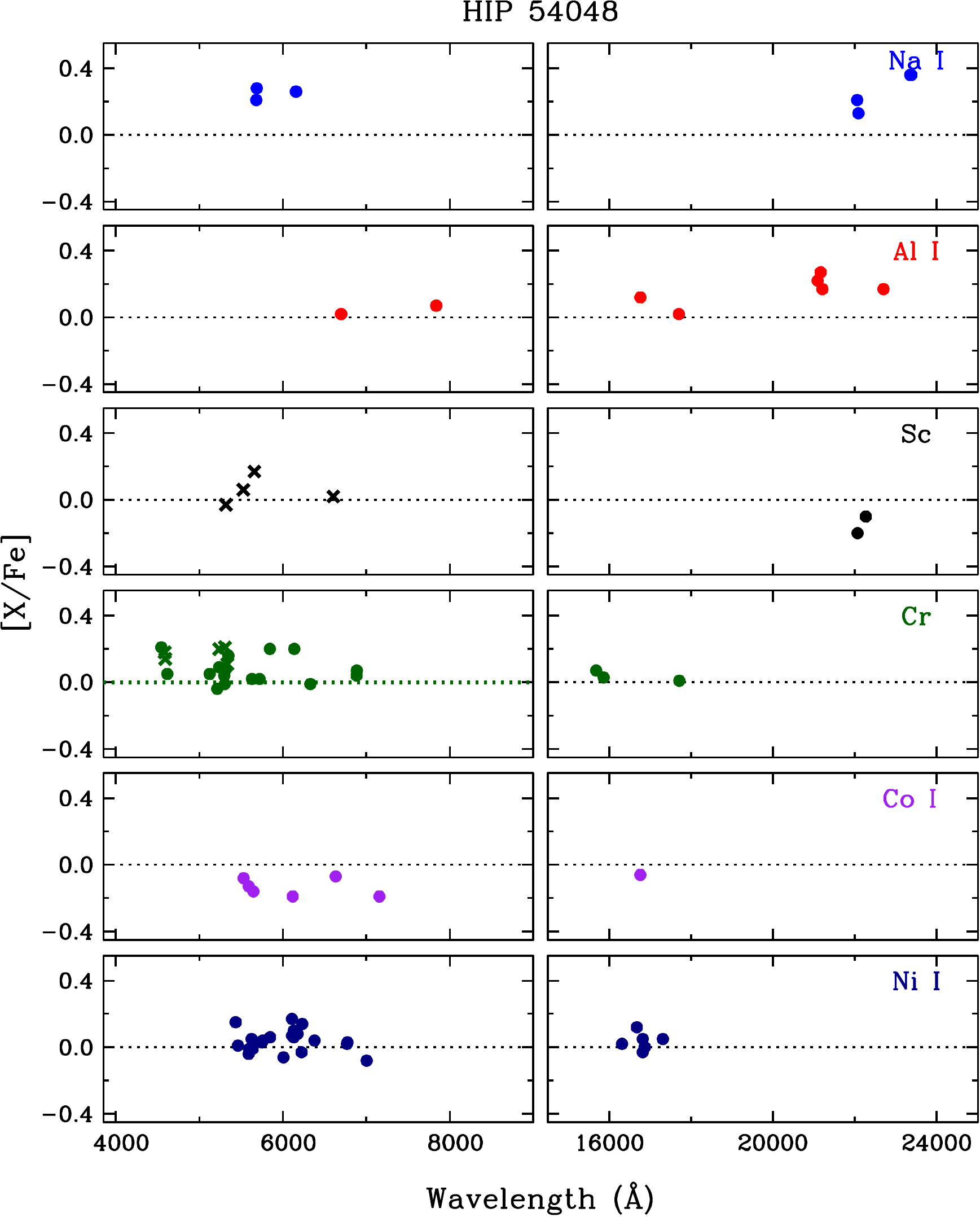}
\caption{
\label{other_elem}
\footnotesize
Relative abundance ratios [X/Fe] of other elements in \ffor.
Sc and Cr panels include both neutral (dots) and ionized (crosses) species 
of the same element.
}
\end{figure}

\begin{figure}
\epsscale{0.85}
\plotone{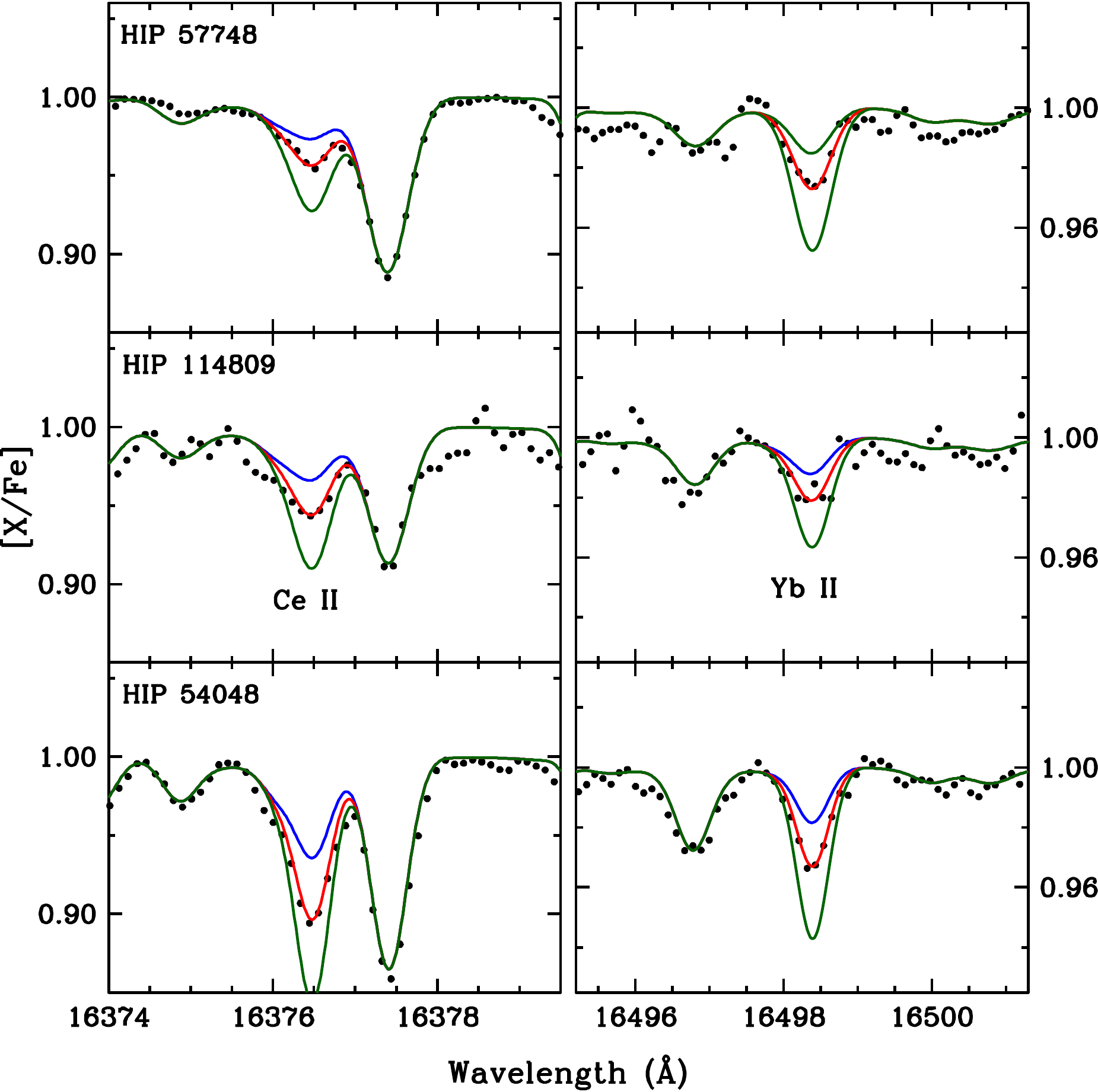}
\caption{
\label{ncapts}
\footnotesize
Neutron capture elements \species{Ce}{ii} (left panel) and \species{Yb}{ii} (right panel) detected in our
stars. The syntheses illustrated with blue and green lines depart 
from the best fit (red line) by $\pm$0.3 dex. 
The lines, colors and symbols are the same as in Figure~\ref{carbiso}.
}
\end{figure}

\begin{figure}
\epsscale{0.6}
\plotone{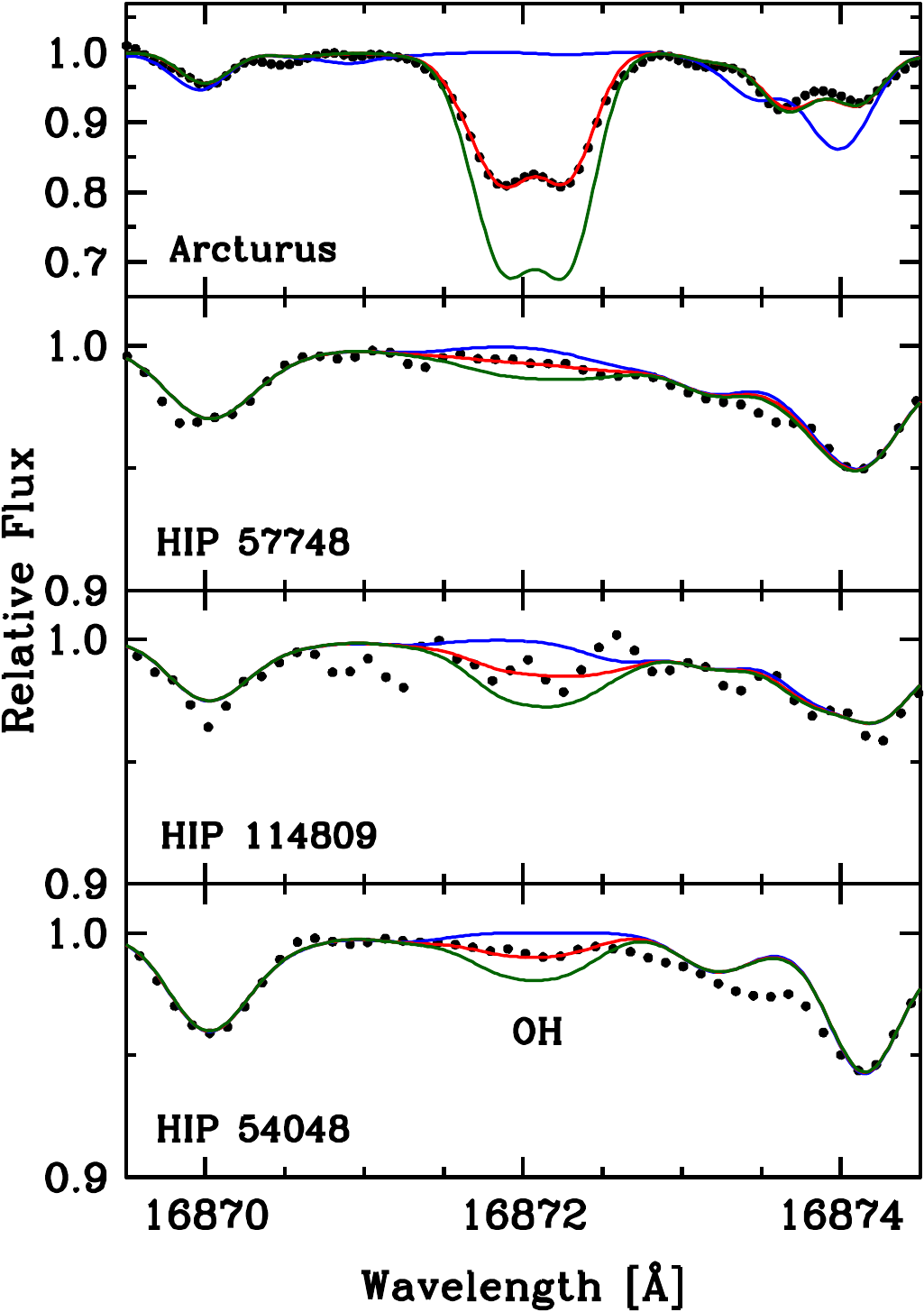}
\caption{
\label{OH}
\footnotesize
OH line at 16872 \AA\ in all RHB stars.  
Since the OH lines are considerably weak in all RHB stars, the syntheses shown 
with blue lines are assumed to be no OH detection.
The syntheses illustrated with green line depart from the best fit (red line) by $+$0.3 dex. 
The lines, colors and symbols are the same as in Figure~\ref{carbiso}.
}
\end{figure}

\begin{figure}
\epsscale{0.95}
\plotone{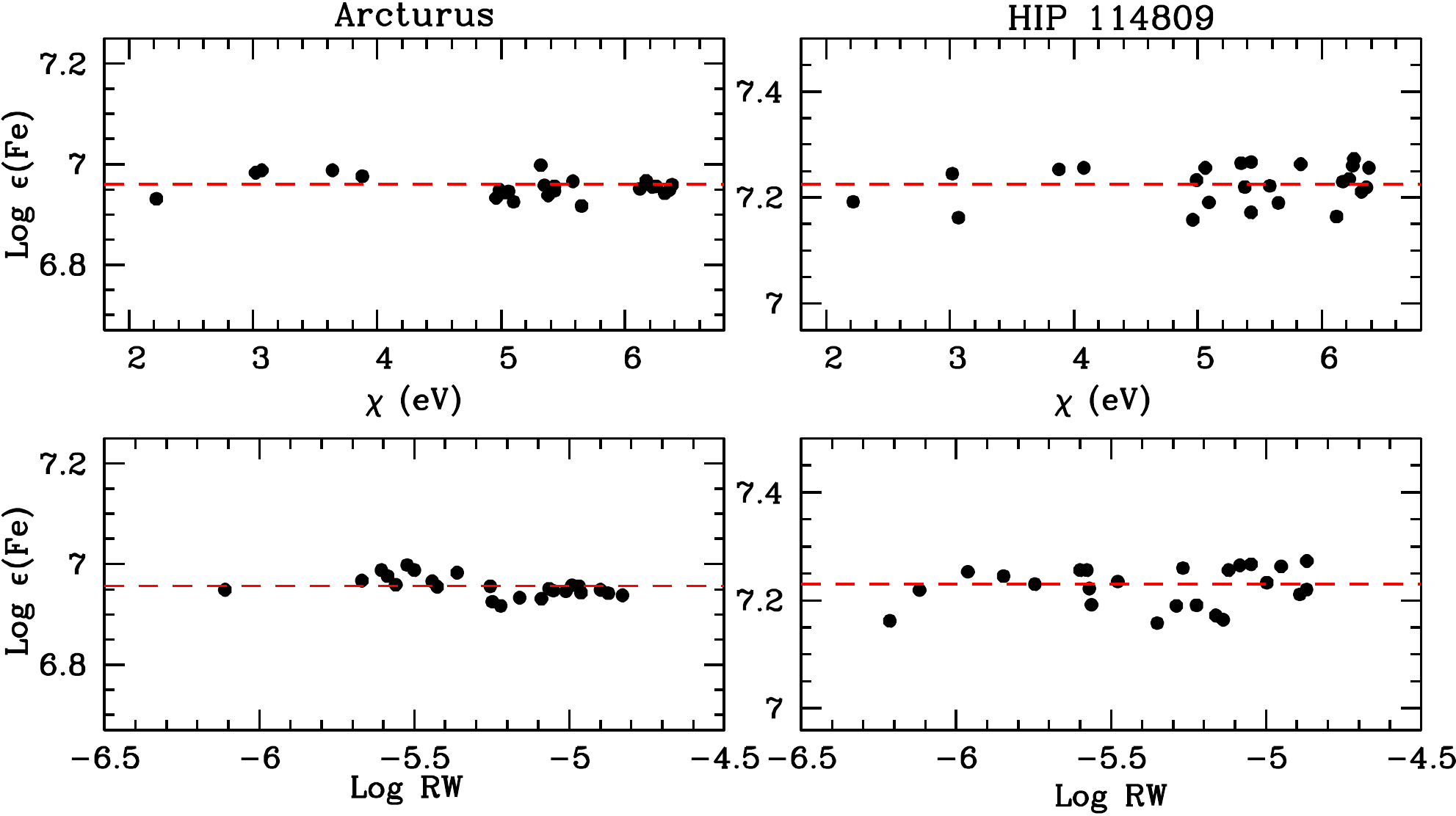}
\caption{
\label{FeTeff}
\footnotesize
Testing the EW measurements of \species{Fe}{i} lines located in the $H$ and $K$ bands 
by setting the model atmosphere parameters obtained from the optical lines
(\S\ref{model}). Top panel shows the line abundances plotted as a function of the excitation 
potential $\chi$ (eV), and the bottom panel as a function of the reduced width RW.
}
\end{figure}

Phosphorus is one of the fundamental building blocks 
of life but its nucleosynthetic origin is still under discussion. 
Recent studies point to massive star explosions (hypernova, 
\eg, \citealt{kobayashi06,cescutti12}).
There are only two useful \species{P}{i} lines in the $UV$ and 
two in the $z$ band.
The near-$IR$ \species{P}{i} lines located in the $H$ band,
15711.5 and 16482.9 \AA\ (Figure~\ref{PK}) are weak but our high-S/N data 
yield detections. We estimated P abundances around solar for our RHB stars.

K is produced mainly by oxygen burning and its production rate during 
nucleosynthesis is still under debate. 
For example, according to \cite{timmes95} [K/Fe] ratios decrease towards 
lower metallicities, while \cite{samland98} and \cite{goswami00} suggest 
[K/Fe] decline in the disk and super solar ratios in halo stars. 
There are observational studies that favor the 
theoretical predictions of \cite{samland98} and \cite{goswami00}, 
e.g., \cite{gratton87,chen00,takeda02,zhang06,takeda09,mucciarelli17}. 
However, the analytical difficulties of the 
\species{K}{i} resonance lines render them problematical to map the 
behavior of K in the Galaxy. 
The $H$ band spectral region provides two more \species{K}{i} lines
(Figure~\ref{PK}). 
Recently, \cite{hawkins16} investigated more than 20 elements 
using a massive Galactic sample from APOGEE data, including [K/Fe]$-$[Fe/H]
relation for $-1.0$~$\lesssim$~[Fe/H]~$<$~0.5 (Figure 16 in 
\citeauthor{hawkins16}). 
Their result also supports the predictions of  \citeauthor{samland98} and 
\citeauthor{goswami00}.
With higher IGRINS $H$ band resolution, we were able to study 15163.1 and 
15168.4 \AA\ \species{K}{i} lines in more detail. 
The K abundances we derived from these lines are around solar in our stars.

Our IGRINS abundance analysis of Fe-group elements 
Sc, Ti, Cr, Co and Ni has been highlighted by consideration of the
substantial isotopic substructure of $IR$ \species{Ti}{i} lines.
Accounting for the isotopic effects yields good agreement with observed
\species{Ti}{i} line profiles in the Sun, Arcturus, and our three RHB stars.
The general agreement between optical and $IR$ abundances for the Fe-peak 
elements is good with the exception of Sc. 
Unfortunately, the optical and $IR$ Sc abundances are based on different
species of this element and the cause of the small ($\sim$0.15~dex) 
offset between the Sc abundances in the two wavelength regions cannot be
traced easily.

The presence of \ncap\ elements in the near-$IR$ were previously discussed by
e.g. \cite{hawkins16,hassel16,cunha17}. 
We determined the abundances of three \ncap\ 
elements in the near-$IR$, all located in the $H$ band. 
Among them, Yb is important because its optical counterpart at 3694 \AA\ 
falls into a very crowded and low-flux spectral region. 
We derived near-solar abundances for Yb.  
The Ce and Nd abundances from both wavelength regions show substantial
star-to-star scatter, and needs further investigation. 

We computed the CNO abundances using CO, CN and OH 
molecular features that are present in the near-$IR$ at considerable amounts. 
Our mildly metal-poor stars are relatively hot for OH molecular lines 
to survive; we could only detect five useful ones in our RHB stars.  
The $IR$ O abundances are in good agreement with their optical 
counterparts that were obtained from the
forbidden [\species{O}{i}] at 6300 and 6363 \AA.
The C abundances from $K$ band CO lines agree well with the C abundances
derived from the CH and C$_{2}$ molecular lines in the optical.
Finally, we used CN molecular lines in the $H$ band to determine N abundances.
These results are in good accord with ones obtained from optical CN features.

We also measured the C abundances from seven 
high-excitation potential \species{C}{i} lines: three in the optical, and 
four in the near-$IR$. 
As is seen in Table~\ref{tab-cmean}, on average the near-$IR$ 
high-excitation \species{C}{i} lines yield C abundances in reasonable 
agreement with the ones obtained from their optical counterparts.
The mean C abundance from all species in all stars has small internal scatter,
$\sigma$~$\leq$~0.10~dex, and these means are always within 0.02~dex
with those obtained just from the CO lines. 
Since  CH, C$_2$, \species{C}{i}, and CO species have different sensitivities 
to adopted \teff,\ \logg, and O abundance values, their reasonable agreement
in our three program stars suggests that the derived mean C abundances are
robust.
However, this assumption is only based on five stars (including Arcturus 
and the Sun). 
Clearly, further investigation is needed with a larger sample to clarify 
this issue. 
To the best of our knowledge, this study is the first among others
that investigates the C abundances in a more complementary way using four 
different C features from the optical and near-$IR$ wavelength regions.

CNO abundances are the most important indicators of internal 
nucleosynthesis and envelope mixing in evolved stars.
Convective envelope mixing brings up the CN-cycle nuclear 
processed material to the stellar surface. 
As a result, C abundance drops, as the N abundance increases 
(e.g. \citealt{iben64,iben67a,iben67b}). 
One of the most important indicators of the convective mixing is the carbon 
isotopic ratio: \carbiso. 
It decreases from its solar value, $\sim$90, to 
$\sim$20$-$30 during the giant branch evolution 
(e.g. \citealt{charbon94,charbon98,gratton00}), sometimes 
even approaching the CN-cycle equilibrium value of \carbiso = 3$-$4
(e.g. \citealt{caughlan65,sneden86,cottrell86,gratton00}).
Determination of the \carbiso\ ratio in the optical is usually possible from the
triplet of $^{13}$CN lines near 8004.7 \AA. It is the only reliable
feature in the optical region and often really weak.
IGRINS $K$ band has two $\Delta$$v$~=~2 transitions of the $^{12}$CO 
first overtone band heads: (2$-$0) and (3$-$1). The $^{13}$CO (2$-$0) and 
$^{13}$CO (3$-$1) band heads that accompany $^{12}$CO are located near 23440 and
23730 \AA, respectively. We used both of these regions and re-determined 
the \carbiso\ ratios for our stars: 12, 8.5 and 15.5 for \ffor, \fsev\ and \oneone,
respectively (Table~\ref{tab-carbiso}).

The mean C and N abundances from the optical and near-$IR$ for our stars are: [C/Fe] = $-$0.57, [N/Fe] = $+$0.80
for \ffor; [C/Fe] = $-$0.18, [N/Fe] = $+$0.39 for \fsev; and [C/Fe] = $-$0.31, [N/Fe] = $+$0.58
for \oneone. The mean O abundances are within $\pm$~0.1 dex in \ffor\ and \fsev, and slightly 
above solar, [O/Fe] = 0.13 dex, in \oneone. The C abundance is considerably low while the N abundance is substantially
high in \ffor\ compared to other RHB stars. In fact the [C/N] = $-$1.37 of \ffor\ puts its location beyond the [C/N]
limits recently studied by, e.g., \cite{lagarde17,lagarde18}, which might imply a very high initial mass for this star.
Further investigation is needed for \ffor. The [C/N] values for \fsev\ and \oneone\ are $-$0.57 and $-$0.89, respectively. 
These values also indicates high initial masses according to \cite{lagarde17} and \cite{lagarde18}.
In all three cases a thermohaline mixing process is required in order to explain the low \carbiso\ ratios
observed in our stars (Table~\ref{tab-carbiso}). 

Near-$IR$ IGRINS high-resolution spectra provide 
an important new opportunity to study stellar chemical compositions in detail. 
As discussed above, $H$ and $K$ bands often yield more reliable 
abundance results, while providing new features that come with open
questions. 
Our next paper with about 70 RHB stars will explore the issues that 
need further investigation with a larger sample and provide more realistic
statistical results for many elements in the near-$IR$ spectral range.

\acknowledgments
We like to thank Nils Ryde, Brian Thorsbro and Elisabetta Caffau for helpful
discussions. Our work has been supported by The Scientific and Technological
Research Council of Turkey (T\"{U}B\.{I}TAK, project No. 112T929), by the US
National Science Foundation grants AST~1211585 and AST~1616040, and by the
University of Texas Rex G. Baker, Jr. Centennial Research Endowment.
This research has made use of the SIMBAD database, operated at CDS, 
Strasbourg, France.
One of the authors (JEL) is supported by the NSF under Grant AST-1516182.
This work used the Immersion Grating Infrared Spectrometer (IGRINS) that was developed under a collaboration between the University of 
Texas at Austin and the Korea Astronomy and Space Science Institute (KASI) with the financial support of the US National Science Foundation 
under grant AST-1229522, of the University of Texas at Austin, and of the Korean GMT Project of KASI. 

\software{IRAF (\citealt{tody93} and references therein),
          MOOG \citep{sneden73},
          ATLAS \citep{kurucz11}}

\bibliographystyle{apj}
\bibliography{rhbbib} 


\clearpage

\clearpage

\clearpage

\clearpage

\clearpage

\clearpage

\clearpage

\clearpage

\clearpage

\clearpage

\clearpage

\clearpage

\clearpage


\clearpage
\begin{center}
\begin{deluxetable}{ccccccccccc}
\tabletypesize{\footnotesize}
\tablewidth{0pt}
\tablecaption{Basic Program Star Data\label{tab-basic}}
\tablecolumns{9}
\tablehead{
\colhead{Star}                          &
\colhead{Other}                         &
\colhead{$V$\tablenotemark{a}}          &
\colhead{$H$\tablenotemark{a}}        &
\colhead{$K$\tablenotemark{a}}        &
\colhead{$\pi$$_{\rm{Hip}}$\tablenotemark{b}}    &
\colhead{$\pi$$_{\rm{Gaia}}$\tablenotemark{c}}   &
\colhead{$M_V$\tablenotemark{d}}        &
\colhead{Date (UT)\tablenotemark{e}}        &
\colhead{Exposure}        &
\colhead{S/N}        \\
\colhead{}                              &
\colhead{}                              &
\colhead{}                              &
\colhead{}                              &
\colhead{}                              &
\colhead{mas}                       &
\colhead{mas}                       &
\colhead{}                              &
\colhead{(2014)}                    &
\colhead{(s)}                          &
\colhead{}
}
\startdata
HIP 54048  & HD 95870  & 6.34 & 4.45 &  4.32 & 6.04 & 5.84 & 0.18 & May 24 & 480 & $\sim$450  \\
HIP 57748  & HD 102857 & 7.90 & 6.12 & 6.01 & 4.47 & 4.77 & 1.29 & May 27 & 720 & $\sim$400 \\
HIP 114809 & HD 219418 & 6.81 & 4.88 & 4.76 & 6.22 & 4.78 & 0.21 & October 19 & 120 & $\sim$200  \\
\enddata

\tablenotetext{a}{SIMBAD database}
\tablenotetext{b}{\cite{vanleeuwen07}}
\tablenotetext{c}{From GAIA DR2 (\citealt{GAIA16,GAIA18b})}
\tablenotetext{d}{Based on GAIA DR2}
\tablenotetext{e}{Date of IGRINS observations}

\end{deluxetable}
\end{center}

\begin{center}
\begin{deluxetable}{ccccc}
\tabletypesize{\footnotesize}
\tablewidth{0pt}
\tablecaption{Model Atmosphere Parameters\tablenotemark{a}\label{tab-model}}
\tablecolumns{5}
\tablehead{
\colhead{Star}                          &
\colhead{\teff}                         &
\colhead{\logg}                         &
\colhead{[M/H]}                         &
\colhead{\vmicro}                       \\
\colhead{}                              &
\colhead{(K)}                             &
\colhead{}                              &
\colhead{}                              &
\colhead{(km s$^{-1}$)}                   
}
\startdata
HIP 54048  & 5099 & 2.63 & $-$0.16 & 1.38 \\
HIP 57748  & 5307 & 2.34 & $-$0.17 & 1.82 \\
HIP 114809 & 5139 & 2.59 & $-$0.38 & 1.31 \\
\enddata

\tablenotetext{a}{\cite{afsar18a}}

\end{deluxetable}
\end{center}

\clearpage
\begin{center}
\begin{deluxetable}{lrrrrrrrr}
\tabletypesize{\footnotesize}
\tablewidth{0pt}
\tablecaption{Abundances from Individual Transition\label{tab-lines}}
\tablecolumns{9}
\tablehead{
\colhead{Species}                       &
\colhead{$\lambda$}                     &
\colhead{$\chi$}                        &
\colhead{log $gf$}                      &
\colhead{source\tablenotemark{a}}       &
\colhead{method\tablenotemark{b}}       &
\colhead{}                &
\colhead{log $\epsilon$(X)}                &
\colhead{}                \\
\colhead{(X)}                              &
\colhead{(\AA)}                           &
\colhead{(eV)}                            &
\colhead{}                              &
\colhead{}                              &
\colhead{}                              &
\colhead{HIP 54048}                      &
\colhead{HIP 57748}                      &
\colhead{HIP 114809}                     
}
\startdata
\mbox{C I}       &     5052.149 &        7.69 &      -1.30 &     1         &        SYN   &       7.70   &     8.12  &      7.75 \\
\mbox{C I}       &     5380.331 &        7.69 &       -1.62 &    1         &        SYN   &       7.77   &     8.18  &      7.91 \\
\mbox{C I}       &     8335.144 &        7.69 &       -0.44 &    1         &        SYN   &       7.82   &     8.35  &      7.91 \\
\mbox{C I}       &    16021.700 &       9.63 &        0.08 &    3         &        SYN   &       7.73   &     8.14  &       7.86 \\
\mbox{C I}       &    16890.386 &       9.00 &        0.42 &    3         &        SYN   &       7.83   &     8.28  &       7.94 \\
\mbox{C I}       &    17455.979 &       9.69 &        0.16 &    3         &        SYN   &       7.83   &     8.24  &       7.94 \\
\mbox{C I}       &    21023.131 &       9.17 &       -0.40 &    1         &        SYN   &       7.73   &     8.16  &       7.84 \\
\mbox{Na I}     &     5682.633 &        2.10 &       -0.71 &    1         &        SYN   &       6.28   &     6.10  &        6.27 \\
\mbox{Na I}     &     5688.194 &        2.10 &       -1.41 &    1         &        SYN   &       6.35   &     6.10  &        6.27 \\
\mbox{Na I}     &     6154.226 &        2.10 &       -1.56 &    1         &        SYN   &       6.33   &     6.10  &        6.20 \\
... \\
\enddata

\tablenotetext{a}{Sources for the $gf$ values: 1. \cite{kramida14}; 2. \cite{kurucz11}; 3. Revsol: reverse solar analysis;
4. \cite{lobel11}; 5. \cite{ryab15}; 6. \cite{ivans06}; 7. \cite{for10}; 8. \cite{ishi12}; 9. \cite{barbuy07}; 10. \cite{reddy12}
11. \cite{carrera11}; 12. \cite{smith81}, 13. \cite{hamdani00}; 14. \cite{simrag81}; 15. \cite{lawler13}; 16. \cite{wood13}; 
17. \cite{wood14c}; 18. \cite{sobeck07}; 19. \cite{cohen04}; 20. \cite{sobeck11}; 21. \cite{wood14b}; 
22. \cite{jacobson07}; 23. \cite{wiese80}; 24. \cite{lawler09}; 25. \cite{cunha17}; 26. \cite{denhartog03};
27. \cite{hassel16}; 28. \cite{lawler89}; 29. \cite{pehlivan15}; 30.  \cite{lawler15}.  }
\tablenotetext{b}{SYN = spectrum synthesis, EW = equivalent width matching.}

\tablecomments{This table is available in its entirety in a machine-readable form in the online journal.}

\end{deluxetable}
\end{center}

\clearpage
\begin{center}
\begin{deluxetable}{lrrrrrr}
\tabletypesize{\footnotesize}
\tablewidth{0pt}
\tablecaption{HIP 54048 Mean Abundances\label{tab-54048}}
\tablecolumns{7}
\tablehead{
\colhead{Species}                       &
\colhead{[X/Fe]}                         &
\colhead{$\sigma$}                      &
\colhead{\#}                            &
\colhead{[X/Fe]}                         &
\colhead{$\sigma$}                      &
\colhead{\#}                            \\
\colhead{}                              &
\colhead{optical}                              &
\colhead{}                       &
\colhead{}                              &
\colhead{infrared}                              &
\colhead{}                      &
\colhead{}
}
\startdata
\mbox{C}\tablenotemark{a}	&	$-$0.62	&	0.01	&	\nodata	&	$-$0.53	&	0.00	&	\nodata	\\
\mbox{N}\tablenotemark{b}	&	0.79	&		&	\nodata	&	0.81	&	0.05	&	\nodata	\\
\mbox{O}\tablenotemark{c}	&	$-$0.07	&	0.12	&	2	&	$-$0.06	&	0.06	&	8	\\
\mbox{Na I}	&	0.25	&	0.03	&	4	&	0.27	&	0.11	&	4	\\
\mbox{Mg I}	&	0.14	&	0.02	&	2	&	0.07	&	0.03	&	11	\\
\mbox{Al I}	&	0.05	&	0.03	&	4	&	0.16	&	0.09	&	6	\\
\mbox{Si I}	&	0.24	&	0.06	&	16	&	0.22	&	0.04	&	12	\\
\mbox{P I}  	&	\nodata	&	\nodata	&	\nodata	&	0.04	&	0.00	&	2	\\
\mbox{S I}	        &	0.01	&	\nodata	&	1	&	0.04	&	0.02	&	10	\\
\mbox{K I} 	&	0.61	&	\nodata	&	1	&	0.06	&	0.10	&	2	\\
\mbox{Ca I}	&	0.16	&	0.04	&	10	&	0.20	&	0.03	&	11	\\
\mbox{Sc I}	&	\nodata	&	\nodata	&	\nodata	&	$-$0.15	&	0.07	&	2	\\
\mbox{Sc II}	&	0.07	&	0.07	&	6	&	\nodata	&	\nodata	&	\nodata	\\
\mbox{Ti I}	&	$-$0.02	&	0.05	&	10	&	0.02	&	0.02	&	10	\\
\mbox{Ti II}	&	0.07	&	0.07	&	6	&	$-$0.08	&	\nodata	&	1	\\
\mbox{Cr I}	&	0.07	&	0.08	&	16	&	0.04	&	0.03	&	3	\\
\mbox{Cr II}	&	0.17	&	0.04	&	5	&	\nodata	&	\nodata	&	\nodata	\\
\mbox{Fe I}\tablenotemark{d}	&	7.33	&	0.06	&	68	&	7.39	&	0.04	&	26	\\
\mbox{Fe II}\tablenotemark{d}	&	7.33	&	0.05	&	12	&	\nodata	&	\nodata	&	\nodata	\\
\mbox{Co I}	&	$-$0.14	&	0.05	&	6	&	$-$0.06	&	\nodata	&	1	\\
\mbox{Ni I}	&	0.04	&	0.06	&	22	&	0.04	&	0.05	&	6	\\
\mbox{Ce II}	&	0.16	&	0.04	&	4	&	0.29	&	0.07	&	6	\\
\mbox{Nd II}	&	0.25	&	0.04	&	2	&	0.37	&	\nodata	&	1	\\
\mbox{Yb II}	&	\nodata	&	\nodata	&	\nodata	&	0.13	&	\nodata	&	1	\\                    
\enddata

\tablenotetext{a}{C abundance: optical from CH, C$_2$, infrared from CO}
\tablenotetext{b}{N abundance: optical and infrared from CN}
\tablenotetext{c}{O abundance: optical from [O I], infrared from OH}
\tablenotetext{d}{\logeps\ values are given for \species{Fe}{i} and \species{Fe}{ii}} 

\end{deluxetable}
\end{center}

\clearpage
\begin{center}
\begin{deluxetable}{lrrrrrr}
\tabletypesize{\footnotesize}
\tablewidth{0pt}
\tablecaption{HIP 57748 Mean Abundances\label{tab-57748}}
\tablecolumns{7}
\tablehead{
\colhead{Species}                       &
\colhead{[X/Fe]}                         &
\colhead{$\sigma$}                      &
\colhead{\#}                            &
\colhead{[X/Fe]}                         &
\colhead{$\sigma$}                      &
\colhead{\#}                            \\
\colhead{}                              &
\colhead{}                              &
\colhead{optical}                       &
\colhead{}                              &
\colhead{}                              &
\colhead{infrared}                      &
\colhead{}
}
\startdata
\mbox{C}\tablenotemark{a}	&	-0.24	&	0.06	&	\nodata	&	-0.12	&	0.04	&	\nodata	\\
\mbox{N}\tablenotemark{b}	&	0.40	&	\nodata	&	\nodata	&	0.37	&	0.02	&	\nodata	\\
\mbox{O}\tablenotemark{c}	&	-0.07	&	0.06	&	2	&	0.03	&	0.15	&	\nodata	\\
\mbox{Na I}	&	0.03	&	\nodata	&	4	&	-0.05	&	0.10	&	4	\\
\mbox{Mg I}	&	0.18	&	0.04	&	2	&	0.15	&	0.09	&	11	\\
\mbox{Al I}	&	-0.15	&	0.01	&	4	&	-0.06	&	0.11	&	6	\\
\mbox{Si I}	&	0.28	&	0.07	&	13	&	0.26	&	0.04	&	11	\\
\mbox{P I}	&	\nodata	&	\nodata	&	\nodata	&	0.03	&	0.04	&	2	\\
\mbox{S I}	&	0.07	&	\nodata	&	1	&	0.10	&	0.04	&	10	\\
\mbox{K I}	&	0.62	&	\nodata	&	1	&	-0.03	&	0.00	&	2	\\
\mbox{Ca I}	&	0.03	&	0.05	&	11	&	0.05	&	0.05	&	11	\\
\mbox{Sc I}	&	\nodata	&	\nodata	&	\nodata	&	\nodata	&	\nodata	&	\nodata	\\
\mbox{Sc II}	&	-0.22	&	0.06	&	5	&	\nodata	&	\nodata	&	\nodata	\\
\mbox{Ti I}	&	-0.19	&	0.06	&	8	&	-0.10	&	0.03	&	9	\\
\mbox{Ti II}	&	-0.07	&	0.06	&	5	&	-0.24	&	\nodata	&	1	\\
\mbox{Cr I}	&	0.00	&	0.06	&	16	&	-0.01	&	0.03	&	3	\\
\mbox{Cr II}	&	0.16	&	0.05	&	4	&	\nodata	&	\nodata	&	\nodata	\\
\mbox{Fe I}\tablenotemark{d}	&	7.33	&	0.06	&	67	&	7.40	&	0.05	&	25	\\
\mbox{Fe II}\tablenotemark{d}	&	7.31	&	0.05	&	12	&	\nodata	&	\nodata	&	\nodata	\\
\mbox{Co I}	&	-0.08	&	0.04	&	5	&	-0.08	&	\nodata	&	1	\\
\mbox{Ni I}	&	0.04	&	0.06	&	24	&	0.03	&	0.07	&	6	\\
\mbox{Ce II}	&	-0.23	&	0.04	&	4	&	-0.26	&	0.02	&	3	\\
\mbox{Nd II}	&	-0.35	&	0.04	&	2	&	-0.15	&	\nodata	&	1	\\
\mbox{Yb II}	&	\nodata	&	\nodata	&	\nodata	&	-0.01	&	\nodata	&	1	\\
\enddata

\tablenotetext{a}{C abundance: optical from CH, C$_2$, infrared from CO}
\tablenotetext{b}{N abundance: optical and infrared from CN}
\tablenotetext{c}{O abundance: optical from [O I], infrared from OH}
\tablenotetext{d}{\logeps\ values are given for \species{Fe}{i} and \species{Fe}{ii}} 

\end{deluxetable}
\end{center}

\clearpage
\begin{center}
\begin{deluxetable}{lrrrrrr}
\tabletypesize{\footnotesize}
\tablewidth{0pt}
\tablecaption{HIP 114809 Mean Abundances\label{tab-114809}}
\tablecolumns{7}
\tablehead{
\colhead{Species}                       &
\colhead{[X/Fe]}                         &
\colhead{$\sigma$}                      &
\colhead{\#}                            &
\colhead{[X/Fe]}                         &
\colhead{$\sigma$}                      &
\colhead{\#}                            \\
\colhead{}                              &
\colhead{}                              &
\colhead{optical}                       &
\colhead{}                              &
\colhead{}                              &
\colhead{infrared}                      &
\colhead{}
}
\startdata
\mbox{C}\tablenotemark{a}	&	$-$0.35	&	0.04	&	\nodata	&	$-$0.26	&	0.02	&	\nodata	\\
\mbox{N}\tablenotemark{b}	&	0.57	&	\nodata	&	\nodata	&	0.58	&	0.02	&	\nodata	\\
\mbox{O}\tablenotemark{c}	&	0.11	&	0.04	&	2	&	0.15	&	0.04	&	\nodata	\\
\mbox{Na I}	&	0.34	&	0.03	&	4	&	0.36	&	0.05	&	4	\\
\mbox{Mg I}	&	0.16	&	0.04	&	2	&	0.14	&	0.04	&	11	\\
\mbox{Al I}	&	0.10	&	0.05	&	4	&	0.20	&	0.09	&	6	\\
\mbox{Si I}	&	0.22	&	0.08	&	14	&	0.23	&	0.04	&	11	\\
\mbox{P I}	&	\nodata	&	\nodata	&	\nodata	&	$-$0.03	&	\nodata	&	2	\\
\mbox{S I}	&	0.10	&	\nodata	&	1	&	0.08	&	0.04	&	10	\\
\mbox{K I}	&	0.75	&	\nodata	&	1	&	0.11	&	0.02	&	2	\\
\mbox{Ca I}	&	0.17	&	0.06	&	11	&	0.15	&	0.03	&	11	\\
\mbox{Sc I}	&	\nodata	&	\nodata	&	\nodata	&	\nodata	&	\nodata	&	\nodata	\\
\mbox{Sc II}	&	0.19	&	0.07	&	6	&	\nodata	&	\nodata	&	\nodata	\\
\mbox{Ti I}	&	0.01	&	0.08	&	9	&	0.07	&	0.03	&	7	\\
\mbox{Ti II}	&	0.16	&	0.07	&	5	&	0.00	&	\nodata	&	1	\\
\mbox{Cr I}	&	0.04	&	0.08	&	11	&	0.03	&	0.04	&	3	\\
\mbox{Cr II}	&	0.17	&	0.10	&	5	&	\nodata	&	\nodata	&	\nodata	\\
\mbox{Fe I}\tablenotemark{d}	&	7.17	&	0.06	&	64	&	7.18	&	0.04	&	24	\\
\mbox{Fe II}\tablenotemark{d}	&	7.12	&	0.05	&	12	&	\nodata	&	\nodata	&	\nodata	\\
\mbox{Co I}	&	$-$0.05	&	0.03	&	5	&	$-$0.05	&	\nodata	&	1	\\
\mbox{Ni I}	&	0.10	&	0.07	&	23	&	0.12	&	0.06	&	6	\\
\mbox{Ce II}	&	0.01	&	0.02	&	3	&	0.09	&	0.09	&	5	\\
\mbox{Nd II}	&	0.08	&	0.06	&	2	&	\nodata	&	\nodata	&	1	\\
\mbox{Yb II}	&	\nodata	&	\nodata	&	\nodata	&	0.07	&	\nodata	&	1	\\
\enddata

\tablenotetext{a}{C abundance: optical from CH, C$_2$, infrared from CO}
\tablenotetext{b}{N abundance: optical and infrared from CN}
\tablenotetext{c}{O abundance: optical from [O I], infrared from OH}
\tablenotetext{d}{\logeps\ values are given for \species{Fe}{i} and \species{Fe}{ii}} 

\end{deluxetable}
\end{center}

\clearpage
\begin{center}
\begin{deluxetable}{lcccccccccccc}
\tabletypesize{\footnotesize}
\tablewidth{0pt}
\tablecaption{\logeps(C)Values From Individual Features\label{tab-c}}
\tablecolumns{12}
\tablehead{
\colhead{Star}                          &
\colhead{CH$^{a}$}                 &
\colhead{C$_{2}$$^{a}$}    &
\colhead{\species{C}{i}}          &
\colhead{\species{C}{i}}        &
\colhead{\species{C}{i}}        &
\colhead{\species{C}{i}}  &
\colhead{\species{C}{i}}       &
\colhead{\species{C}{i}}        &
\colhead{\species{C}{i}}       &
\colhead{CO}                    &
\colhead{CO}                      \\
\colhead{}                &
\colhead{G-band}     &
\colhead{Swan}        &
\colhead{5052}                              &
\colhead{5380}                              &
\colhead{8335}                              &
\colhead{16022}    &
\colhead{16890}   &
\colhead{17456}     &
\colhead{21023}                              &
\colhead{23400}                          &
\colhead{23700}                \\
\colhead{}                &
\colhead{(\AA)}     &
\colhead{(\AA)}                              &
\colhead{(\AA)}                              &
\colhead{(\AA)}                              &
\colhead{(\AA)}        &
\colhead{(\AA)}    &
\colhead{(\AA)}   &
\colhead{(\AA)}     &
\colhead{(\AA)}                              &
\colhead{(\AA)}                          &
\colhead{(\AA)}     
}
\startdata
\ffor	        &	7.64	&    7.65     &   7.70	&	7.77	&	7.82	&	7.73	&	7.83	&	7.83	&	7.73	&	7.73	 & 7.73  \\
\fsev		&	7.98	&    8.06     &   8.12	&	8.18	&	8.35	&	8.14	&	8.28	&	8.24	&	8.16	&      8.16   &  8.12  \\
\oneone 	&	7.72	&    7.78     &   7.75	&	7.91	&      7.91	&      7.86	&	7.94	&	7.94	&	7.84	&      7.85   &  7.82  \\
\enddata

\tablenotetext{a}{G-band refers to the band heads in the 4300-4330 \AA\ region. Swan refers to the band heads near 5160 and 5631 \AA.}

\end{deluxetable}
\end{center}

\clearpage
\begin{center}
\begin{deluxetable}{lccccccc}
\tabletypesize{\footnotesize}
\tablewidth{0pt}
\tablecaption{Mean \logeps(C) Values of Species and Final Derived Elemental Abundance\label{tab-cmean}}
\tablecolumns{8}
\tablehead{
\colhead{Star}                          &
\colhead{CH}                         &
\colhead{C$_{2}$}    &
\colhead{\species{C}{i}$_{(\rm opt)}$}  &
\colhead{\species{C}{i}$_{(\rm IR)}$}  &
\colhead{CO}                    &
\colhead{$\langle$\logeps(C)$\rangle$}                    &
\colhead{$\sigma$}         
}
\startdata
\ffor	        &	7.64	&	7.65	&	7.76	&	7.78	&	7.73	&	7.71	&	0.06	  \\
\fsev		&	7.98	&	8.06	&	8.22	&	8.21	&	8.14	&	8.12	&	0.10	  \\
\oneone 	&	7.72	&	7.78	&	7.86	&      7.90	&	7.84 &      	7.82  &	0.07	  \\
\enddata

\end{deluxetable}
\end{center}

 \clearpage
\begin{center}
\begin{deluxetable}{lccrccr}
\tabletypesize{\footnotesize}
\tablewidth{0pt}
\tablecaption{Fe Abundances From Common \species{Fe}{i} Lines Used in Both SYN and EW Methods for \oneone.\label{tab-synew}}
\tablecolumns{7}
\tablehead{
\colhead{Species}                       &
\colhead{$\lambda$}                     &
\colhead{$\chi$}                        &
\colhead{log $gf$}                      &
\colhead{log $\epsilon$(Fe)$_{\rm SYN}$}       &
\colhead{log $\epsilon$(Fe)$_{\rm EW}$}      &
\colhead{$\Delta$$_{\small{\rm (SYN-EW)}}$} 
}
\startdata
\mbox{Fe I}	&	15194.49	&	2.22	&	$-$4.75	&	7.09	&	7.18	&	$-$0.09	\\
\mbox{Fe I}	&	15207.53	&	5.39	&	0.08	&	7.24	&	7.22	&	0.02	\\
\mbox{Fe I}	&	15343.79	&	5.65	&	$-$0.69	&	7.14	&	7.18	&	$-$0.04	\\
\mbox{Fe I}	&	15493.52	&	6.36	&	$-$1.06	&	7.19	&	7.17	&	0.02	\\
\mbox{Fe I}	&	15648.51	&	5.43	&	$-$0.70	&	7.14	&	7.17	&	$-$0.03	\\
\mbox{Fe I}	&	15662.01	&	5.83	&	0.07	&	7.17	&	7.26	&	$-$0.09	\\
\mbox{Fe I}	&	15761.31	&	6.25	&	$-$0.16	&	7.17	&	7.26	&	$-$0.09	\\
\mbox{Fe I}	&	15858.66	&	5.58	&	$-$1.25	&	7.14	&	7.21	&	$-$0.07	\\
\mbox{Fe I}	&	15980.73	&	6.26	&	0.72	&	7.19	&	7.27	&	$-$0.08	\\
\mbox{Fe I}	&	16009.61	&	5.43	&	$-$0.55	&	7.18	&	7.27	&	$-$0.09	\\
\mbox{Fe I}	&	16165.03	&	6.32	&	0.75	&	7.24	&	7.21	&	0.03	\\
\mbox{Fe I}	&	16171.93	&	6.38	&	$-$0.51	&	7.22	&	7.25	&	$-$0.03	\\
\mbox{Fe I}	&	17420.83	&	3.88	&	$-$3.52	&	7.24	&	7.24	&	0.00	\\
\mbox{Fe I}	&	21178.16	&	3.02	&	$-$4.24	&	7.14	&	7.24	&	$-$0.10	\\
\mbox{Fe I}	&	21238.47	&	4.96	&	$-$1.37	&	7.14	&	7.15	&	$-$0.01	\\
\mbox{Fe I}	&	21284.35	&	3.07	&	$-$4.51	&	7.14	&	7.16	&	$-$0.02	\\
\mbox{Fe I}	&	21735.46	&	6.18	&	$-$0.73	&	7.21	&	7.23	&	$-$0.02	\\
\mbox{Fe I}	&	22257.11	&	5.06	&	$-$0.82	&	7.24	&	7.25	&	$-$0.01	\\
\mbox{Fe I}	&	22260.18	&	5.09	&	$-$0.98	&	7.17	&	7.19	&	$-$0.02	\\
\mbox{Fe I}	&	22419.98	&	6.22	&	$-$0.30	&	7.19	&	7.23	&	$-$0.04	\\
\mbox{Fe I}	&	22473.26	&	6.12	&	0.32	&	7.22	&	7.16	&	0.06	\\
\mbox{Fe I}	&	22619.84	&	4.99	&	$-$0.51	&	7.17	&	7.23	&	$-$0.06	\\
\mbox{Fe I}	&	23308.48	&	4.08	&	$-$2.73	&	7.17	&	7.25	&	$-$0.08	\\
\enddata

\tablenotetext{}{SYN = spectrum synthesis, EW = equivalent width matching.}

\end{deluxetable}
\end{center}

 \clearpage
\begin{center}
\begin{deluxetable}{lccc}
\tabletypesize{\footnotesize}
\tablewidth{0pt}
\tablecaption{\carbiso\ Ratios of Program Stars From Optical and $IR$ Regions\label{tab-carbiso}}
\tablecolumns{3}
\tablehead{
\colhead{Stars}                       &
\colhead{$^{13}$CN}                              &
\colhead{$^{13}$CO (2$-$0)}           &
 \colhead{$^{13}$CO (3$-$1)}             \\
 \colhead{}                            &
\colhead{(8004 \AA)}                              &
\colhead{(23440 \AA)}           &
 \colhead{(23730 \AA)}                                             
}
\startdata
\ffor	&	10	&	12	& 12  \\
\fsev	&	6	&	 8	&    9  \\
\oneone &  \nodata     &   15    & 16  \\
\enddata

\end{deluxetable}
\end{center}

\end{document}